\documentclass[fleqn,usenatbib]{mnras}

\usepackage{newtxtext,newtxmath}

\usepackage[T1]{fontenc}
\usepackage{ae,aecompl}
\usepackage{multirow}

\DeclareUnicodeCharacter{2212}{-}

\usepackage{graphicx}	
\usepackage{amsmath}	
\usepackage{lipsum}

\title[Combined APOGEE-GALAH catalogues]{Combined APOGEE-GALAH stellar catalogues using the \text{Cannon}}

\author[ Nandakumar et al.]{
Govind Nandakumar$^{1,2}$\thanks{E-mail: govind.nandakumar@anu.edu.au},
Michael R. Hayden$^{3,2}$,
Sanjib Sharma$^{3,2}$,
Sven Buder$^{1,2}$,
Martin Asplund$^{4}$,
\newauthor%
Joss~Bland-Hawthorn$^{3,2}$,
Gayandhi~M.~De~Silva$^{5,6}$,
Valentina~{D'Orazi}$^{7}$,
Ken~C.~Freeman$^{1}$,
Janez~Kos$^{8}$,
\newauthor%
Geraint~F.~Lewis$^{3}$,
Sarah~L.~Martell$^{9,2}$,
Katharine~J.~Schlesinger$^{1}$,
Jane~Lin$^{1,2}$,
Jeffrey~D.~Simpson$^{9,2}$,
\newauthor%
Daniel~B.~Zucker$^{10,6}$,
Toma\v{z}~Zwitter$^{8}$,
Thomas Nordlander$^{1,2}$,
Luca Casagrande$^{1,2}$,
Karin Lind$^{11}$,
\newauthor%
Klemen C\^otar$^{8}$,
Dennis Stello$^{3,9,2}$,
Robert A. Wittenmyer$^{12}$ and
Thor Tepper-Garcia$^{3,2,13}$\\
\\
$^{1}$Research School of Astronomy $\&$ Astrophysics, Australian National University, ACT 2611, Australia\\
$^{2}$Centre of Excellence for Astrophysics in Three Dimensions (ASTRO-3D), Australia\\
$^{3}$Sydney Institute for Astronomy, School of Physics, A28, The University of Sydney, NSW 2006, Australia\\
$^{4}$Max Planck Institute for Astrophysics, Karl-Schwarzschild-Str. 1, D-85741 Garching, Germany\\
$^{5}$Australian Astronomical Optics, Faculty of Science and Engineering, Macquarie University, Macquarie Park, NSW 2113, Australia\\
$^{6}$Macquarie University Research Centre for Astronomy, Astrophysics \& Astrophotonics, Sydney, NSW 2109, Australia\\
$^{7}$Istituto Nazionale di Astrofisica, Osservatorio Astronomico di Padova, vicolo dell'Osservatorio 5, 35122, Padova, Italy\\
$^{8}$Faculty of Mathematics and Physics, University of Ljubljana, Jadranska 19, 1000 Ljubljana, Slovenia\\
$^{9}$School of Physics, UNSW, Sydney, NSW 2052, Australia\\
$^{10}$Department of Physics and Astronomy, Macquarie University, Sydney, NSW 2109, Australia\\
$^{11}$Department of Astronomy, Stockholm University, AlbaNova University Center, SE-106 91 Stockholm - Sweden\\
$^{12}$University of Southern Queensland, Centre for Astrophysics, USQ Toowoomba, QLD 4350 Australia\\
$^{13}$ Centre for Integrated Sustainability Analysis, The University of Sydney, 2006, Australia
}

\date{Accepted XXX. Received YYY; in original form ZZZ}

\pubyear{2020}

\begin{document}
\label{firstpage}
\pagerange{\pageref{firstpage}--\pageref{lastpage}}
\maketitle

\begin{abstract}
APOGEE and GALAH are two high resolution multi-object spectroscopic surveys that provide fundamental stellar parameters and multiple elemental abundance estimates for about half a million stars in the Milky Way. Both surveys observe in different wavelength regimes and use different data reduction pipelines leading to significant offsets and trends in stellar parameters and abundances for the common stars observed in both surveys. Such systematic differences/offsets in stellar parameters and abundances make it difficult to effectively utilise them to investigate Galactic abundance trends in spite of the unique advantage provided by their complementary sky coverage and different Milky Way components they observe. Hence, we use the \textit{Cannon} data-driven method selecting a training set of 4418 common stars observed by both surveys. This enables the construction of two catalogues, one with  the APOGEE scaled and the other with the GALAH scaled stellar parameters. Using repeat observations in APOGEE and GALAH, we find high precision in metallicity ($\sim$ 0.02-0.4 dex) and alpha abundances ($\sim$ 0.02-0.03 dex) for spectra with good signal-to-noise ratio (SNR $>$ 80 for APOGEE, SNR $>$40 for GALAH). We use open and globular clusters to validate our parameter estimates and find small scatter in metallicity (0.06 dex) and alpha abundances (0.03 dex) in APOGEE scaled case. The final catalogues have been cross matched with the \textit{Gaia} EDR3 catalogue to enable their use to carry out detailed chemo-dynamic studies of the Milky Way from perspectives of APOGEE and GALAH.

\end{abstract}




\section{Introduction}
\label{sec:intro} 

  The field of Galactic archaeology \citep{Freeman:2002} deals with dissecting the Milky Way into its various components with the aim to unravel the processes that contributed to the formation and evolution of our Galaxy. Stellar spectroscopy plays a crucial role in this field by enabling accurate measurement of stellar parameters and detailed chemical compositions of stars in the Galaxy. Such measurements are crucial in tracing these stars back to their birth sites providing clues to the physical processes that led them to their present position in the Galaxy.

  There are a plethora of data available in the form of spectra, astrometric and photometric information as well as multi wavelength maps with the advent of large scale spectroscopic (Apache Point Observatory Galactic Evolution Experiment/APOGEE: \citealt{Eisenstein:2011}, RAdial Velocity Experiment/RAVE: \citealt{Steinmetz:2006}, Gaia-ESO: \citealt{Gilmore:2012}, Large Sky Area Multi-Object Fiber Spectroscopic Telescope/LAMOST: \citealt{Cui:2012}, Galactic Archaeology with HERMES/GALAH: \citealt{DeSilva:2015}, Abundances and Radial velocity Galactic Origins Survey/ARGOS: \citealt{Ness:2012}), astrometric (Hipparcos: \citealt{Perryman:1997}, \textit{Gaia}: \citealt{Gaia:2016}) and photometric surveys (Two-Micron All Sky Survey/2MASS: \citealt{Skrutskie:2006}, Sloan Digital Sky Survey/SDSS: \citealt{Stoughton:2002}, Vista Variables in the V\'ia L\'actea/VVV: \citealt{Minniti:2010},the SkyMapper Southern Survey : \citealt{Wolf:2018}). These surveys have enabled the chemo-dynamic characterisation of stellar populations in the Milky way that constitute different Milky Way components like thin disc, thick disc, halo, bulge etc. For example, star count observations in the solar neighborhood (\citealt{Yoshii:1982,Gilmore:1983}) led to the discovery of the thick disc, followed by its characterisation as the old $\alpha$-enhanced population in the double sequence exhibited by the solar neighborhood stars in the [$\alpha$/Fe] vs [Fe/H] plane (\citealt{Fuhrmann:1998,Bensby:2003,Reddy:2006,Adibekyan:2012,Haywood:2013}). At present, data from large scale spectroscopic surveys (\citealt{Anders:2014,Hayden:2015,Weinberg:2019}) have led to the discovery of this trend at different galactocentric radius, R, and average height, $|$Z$|$, across the Galaxy shedding light on the disc formation and evolution scenarios. In addition, many age determination methods have been developed that uses these survey data to provide valuable information about the star formation histories and age metallicity relation of disc stellar populations (\citealt{Casagrande:2011,Bedell:2018,Lin:2020,Nissen:2020}). Secular processes such as radial migration (\citealt{Sellwood:2002,Schonrich:2009,Minchev:2010}) that leads to the mixing of stars across the Galaxy, are also being explored using a combination of accurate phase space information from \textit{Gaia} \citep{GaiaDR2:2018} and chemistry and age information of stars from large scale spectroscopic surveys \citep{Buder:2019}. The discovery of streams and dynamically different stellar populations in the Milky Way halo, considered to be the result of past accretion/merger events (\citealt{Belokurov:2018,Helmi:2018,Ibata:2019,Myeong:2019}) using the \textit{Gaia} data and their further exploration with chemistry from large-scale spectroscopic surveys (Buder et al. in prep) is another example. Multiple components in the Bulge metallicity distribution function (MDF) discovered by multiple individual and large scale spectroscopic observations, are being studied in detail to understand the origin of the Bulge and its connection with the Milky Way bar and Galaxy evolution (\citealt{Ness:2013,Rojas-Arriagada:2017,Rojas-Arriagada:2020}). There are many upcoming surveys (4-metre Multi-Object Spectroscopic Telescope/4MOST : \citet{deJong:2019}, Sloan Digital Sky Survey/SDSS-V : \citet{Kollmeier:2017}, WEAVE : \citet{Dalton:2018}) that will further our understanding of the formation and evolution of the Milky Way and its components.

   The above mentioned large-scale spectroscopic surveys derive fundamental stellar parameters and elemental abundances from observed stellar spectra via dedicated pipelines using spectral fitting routines that fit observed spectra with synthetic spectra generated from stellar model atmospheres, model grids and linelists, all of which are different/specific to the respective survey. Thus, even though there are overlaps in the observed stars between many spectroscopic surveys, there are significant systematic differences in their derived stellar parameters as well as abundances. This difference can also lead to misinterpretation of abundance trends estimated using the derived parameters from different surveys. Thus it is necessary to combine such complementary surveys with their parameters scaled with respect to either survey so that the resulting volume complete sample can be used to map and decipher the global chemo-dynamic trends of stellar populations in the Milky Way.

  One step toward this direction of combining surveys (or scaling them with respect to each other) was made with the introduction of the data driven approach known as the \textit{Cannon} \citep{Ness:2015}. \cite{Ho:2017} used the \textit{Cannon} to derive the stellar parameters for around 450,000 giant stars in LAMOST (low spectral resolution survey) by bringing them to the scale of APOGEE (high spectral resolution) survey. Recently, \cite{Wheeler:2020} used the \textit{Cannon} to estimate abundances representing five different nucleosynthetic channels, for 3.9 million stars in LAMOST, by training LAMOST spectra with GALAH DR2 stellar parameters and abundances (also referred to as "labels") for stars in common to both surveys. The \textit{Cannon} has also been used to propagate information from one survey to another, and to derive higher precision stellar parameters, abundances, mass and age information using survey pipeline estimate as the training set labels (\citealt{Ness:2016,Casey:2017,Buder:2018,Zhang:2019,Hasselquist:2020} etc.). The \textit{Starnet} \citep{Fabbro:2018}, a convolutional neural network model, was able to predict stellar parameters by training on APOGEE spectra with APOGEE Stellar Parameter and Chemical Abundance Pipeline (ASPCAP) labels (T$_\mathrm{eff}$, $\log g$ and [Fe/H]). When compared with the \textit{Cannon} results trained on the same data, the \textit{Starnet} showed similar behaviour, though the \textit{Starnet} performs poorly on small training sets compared to the \textit{Cannon}. A deep neural network designed by \cite{Leung:2019} was used to determine stellar parameters from APOGEE spectra using the full wavelength range, while censored portions of the spectrum were used to derive individual element abundances. The \textit{Payne} \citep{Ting:2019} is another tool that explicitly models spectra as a function of stellar parameters. \cite{Xiang:2019} used the data driven \textit{Payne} to train a model that predicts stellar parameters and abundances for 16 elements from LAMOST DR5 spectra using stars in common with APOGEE DR14 and GALAH DR2 as the training set. Thus there are many tools and methods available to put different surveys on the same scale.

  In this work, we use the data driven approach, the \textit{Cannon} 2 (\citealt{Ness:2015,Casey:2016}), to put the stellar parameters (T$_\mathrm{eff}$, $\log g$ and [Fe/H]) and general [$\alpha$/Fe] abundance on the same scale for the surveys, APOGEE and GALAH. For this, we have to select a training set composed of common stars observed in both the surveys, with high fidelity stellar labels as well as high quality spectra. Since both the surveys yield stellar parameters and abundances with dedicated pipelines from high resolution spectra (though in different wavelength ranges), we cannot choose either survey to be the best. Hence, we carry out the exercise in both ways, i.e., (i) train \textit{Cannon} model on APOGEE spectra with GALAH labels and (ii) train \textit{Cannon} model on GALAH spectra with APOGEE labels and derive stellar parameters and [$\alpha$/Fe] values for both cases. We also (iii) train \textit{Cannon} model on GALAH spectra with GALAH labels and (iv) \textit{Cannon} model on APOGEE spectra with APOGEE labels, so that we can combine (i) and (iii) to provide the GALAH scaled stellar parameter catalogue, and (ii) and (iv) to provide the APOGEE scaled stellar parameter catalogue. 
  
  We describe the data used in this paper in Section \ref{sec:data}. In Section \ref{sec:method}, we give a brief description of the \textit{Cannon}, followed by the training set selection, cross-validation of the \textit{Cannon} estimates compared to the input labels from the training set, testing, flagging and error estimation. We carry out astrophysical validation of our \textit{Cannon} estimates using open and globular cluster members in Section~\ref{sec:valid}. Finally, we discuss the limitations and caveats as well as notable improvements in \textit{Cannon} estimates with respect to survey pipleline estimates in Section~\ref{sec:limitations}

\section{Data}
\label{sec:data}

 We use the latest available data release of APOGEE (DR16) and GALAH (DR3). We make use of the fundamental stellar parameters:  T$_\mathrm{eff}$, $\log g$, [Fe/H] and [$\alpha$/Fe] from those catalogues, along with the stellar spectra for each star.

\subsection{APOGEE}
\label{sec:apogee} 

The Apache Point Observatory Galaxy Evolution Experiment (APOGEE, \citealt{majewski:17})  observes in the near-infrared H-band (15,000--17,000\,\AA), split into 3 bands at high spectral resolution (R $\sim$ 22,500) and high signal-to-noise ratios, S/N, in both the Northern and Southern hemispheres. As part of SDSS-IV \citep{Blanton:2017}, the APOGEE-2 survey makes use of the APOGEE instrument \citep{Wilson:2012} on the Sloan 2.5 m Telescope \citep{Gunn:2006} at the Apache Point Observatory (APO) for Northern hemisphere observations, and the 2.5m du Pont telescope at Las Campanas Observatory (LCO; \citealt{Bowen:1973}) with the twin Near infrared (NIR) spectrograph \citep{Wilson:2019} for Southern hemisphere observations.

We make use of the data products from DR16 \citep{DR16}, which contains a total of 473,307 sources with derived atmospheric parameters and elemental abundances of up to 26 species from the APOGEE Stellar Parameters and Chemical Abundances Pipeline (ASPCAP; \citealt{ASPCAP:2016}) using MARCS grid of atmospheric models \citep{Jonsson:2020}. For our work, we use the calibrated "PARAM" stellar parameters and abundances described in \cite{Jonsson:2020} as they have shown that there are several systematic issues with the spectroscopic (FPARAM) measurements of stellar parameters (Section 6.3 and Figure10 in \citep{Jonsson:2020}).
In addition to the catalogue, we also make use of the "apStar" and "asStar" spectra\footnote{https://data.sdss.org/sas/dr16/apogee/spectro/redux/r12/stars/} which are the combined spectra of multiple visits, all in a common rest frame and identical wavelength solution across all sources.

\subsection{GALAH}
\label{sec:galah} 
Galactic Archaeology with HERMES (GALAH;\citealt{DeSilva:2015}) is a high resolution spectroscopic survey of the Milky Way using the High Efficiency and Resolution Multi-Element Spectrograph (HERMES; \citealt{Barden:2010,Sheinis:2015}) on the Anglo-Australian Telescope. The HERMES spectrograph provides high-resolution (R $\sim$ 28,000) spectra from 4700--7900\,\AA in four wavelength bands.

We make use of data from the latest GALAH data release (GALAH DR3; \citealt{GALAHDR3}), which includes the K2-HERMES survey (\citealt{Wittenmyer:2018, Sharma:2019}) and the TESS-HERMES survey \citep{Sharma:2018}, and provides up to 30 element abundances from various nucleosynthesis channels for 678,423 spectra of 588,571 stars. Here we will refer to this data set collectively as the GALAH survey. Observations are reduced through a standardised pipeline developed for the GALAH survey as described in \cite{Kos:2017}. In this data release, all stellar parameters and elemental abundances have been estimated via the spectrum synthesis code Spectroscopy Made Easy (SME;\citealt{Valenti:1996,Piskunov:2017}) with 1D MARCS stellar atmosphere models \citep{Gustafsson:2008}. In addition to the catalogue, we also make use of the GALAH spectra\footnote{https://docs.datacentral.org.au/galah/dr3/spectra-data-access/}, all in a common rest frame. After removing 12,181 stars which are missing the spectra for some of their chips, we are left with 576,390 GALAH spectra.

\begin{figure*}
	\includegraphics[width=\textwidth]{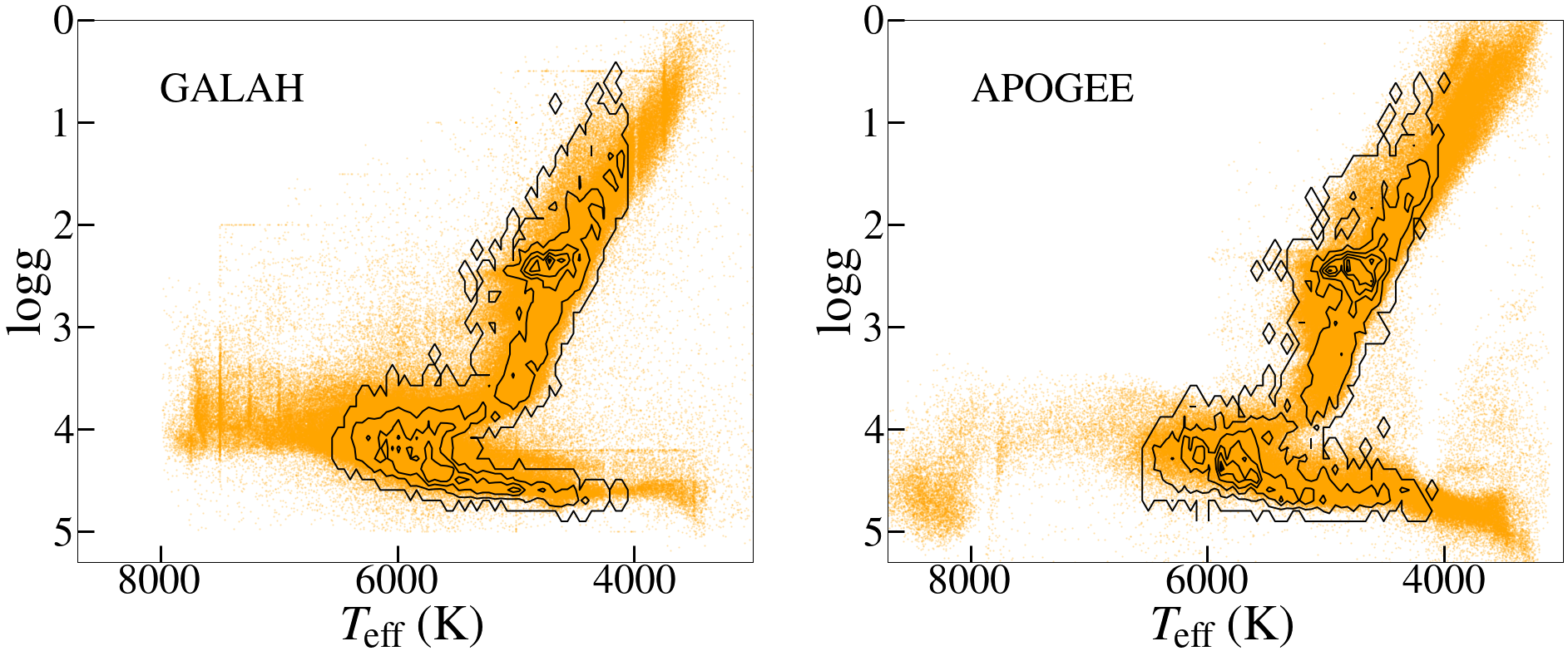}
    \caption{Kiel diagram ($\log g$ vs T$_\mathrm{eff}$) with the respective pipeline estimates for all stars in GALAH (left panel; orange points) and all stars in APOGEE (right panel; orange points) overlaid with the black contours representing the training set which is chosen from among the common stars in both APOGEE and GALAH as described in Section~\ref{sec:trset}.    }
    \label{fig:HR_alphafe}
\end{figure*}

\section{Method}
\label{sec:method}
\subsection{The \textit{Cannon}}
\label{sec:Cannon} 
The \textit{Cannon} \citep{Ness:2015}, in simple terms, is a data driven method which generates a model for stellar spectra based on a training set of spectra, for which the stellar parameters and abundances are known with high fidelity (i.e, high signal-to-noise. ratio (SNR) and/or accurate estimates). This model can then be used to infer the same labels for any set of continuum normalised spectra that have been sampled onto a common wavelength grid with uniform start and end wavelengths as that of the training set. The\textit{ Cannon} relies on the following assumptions: similar labels imply similar spectra and each spectrum is a smooth function of its labels such that changes in labels result in a smooth variation of spectra.

In the training step, the spectral model coefficients are fit at each wavelength pixel while keeping the labels fixed for all training set star spectra. The spectral model thus generated characterises the flux at each wavelength pixel as a function of the given labels, with label coefficient values describing the influence of the corresponding label at a certain wavelength pixel. In the label inference step, the label coefficients are fixed, while likelihood optimization is carried out to predict the labels from the flux values at each wavelength pixel of each test spectrum.

We use the\textit{ Cannon} 2 described in \cite{Casey:2016}\footnote{https://github.com/andycasey/AnniesLasso} and use a quadratic model with T$_\mathrm{eff}$, $\log g$, [Fe/H], [$\alpha$/Fe], microturbulence ($v_{\rm micro}$), and line broadening ($v_{\rm broad}$) as the labels, resulting in a spectral model of the following form :

\begin{equation}
    F_{n\lambda} = v(l_{n}).\theta_{\lambda} + noise
\end{equation}

\noindent where F$_{n\lambda}$ is the flux at each wavelength pixel, $\lambda$, for each star, $n$, in the training set. $\theta_{\lambda}$ is the set of spectral model coefficients for multiple label combinations at each $\lambda$ pixel, $l_{n}$ represents the labels and $v(l_{n})$ is the "vectorizing function", which is in the form of a 2 degree quadratic polynomial and the labels have been normalised or scaled in the following manner:
\begin{equation}
    \hat{l}_{n} = \frac{l_{n} - l_{n,50}}{l_{n,97.5} - l_{n,2.5}}
\end{equation}

\noindent where l$_{n,2.5}$, l$_{n,2.5}$ and l$_{n,2.5}$ are respectively the 2.5$^{th}$, 50$^{th}$ and 97.5$^{th}$ percentile values of the label, $l$, in the training set.

Finally, noise is drawn from a Gaussian distribution with zero mean and variance defined as the sum of the variance in flux, $\sigma_{n\lambda}^{2}$, and excess variance at each wavelength pixel, $s_{\lambda}^{2}$. While variance in flux at each wavelength pixel is obtained from the observed spectra file of each star, excess variance is estimated along with $\theta_{\lambda}$ in the training step by optimising the likelihood function :
\begin{equation}
   \mathrm{arg\,min}_{\theta,s}\left[\sum_{n=0}^{N_\mathrm{stars}-1} \frac{[F_{n\lambda}-v(l_\mathrm{n}).\theta_{\lambda}]^{2}}{\sigma_{n\lambda}^{2} + s_{\lambda}^{2}}+\sum_{n=0}^{N_\mathrm{stars}-1}\mathrm{ln}(\sigma_{n\lambda}^{2} + s_{\lambda}^{2})\right]
\end{equation}\label{Eq:train}


In the test step, hereby called the label inference step, labels corresponding to each test spectrum, m, are predicted by optimising the likelihood function:
\begin{equation}
   \mathrm{arg\,min}_{l}\left[\sum_{\lambda=0}^{N_\mathrm{pix}} \frac{[F_{m\lambda}-v(l_{m}).\theta_{\lambda}]^{2}}{\sigma_{m\lambda}^{2} + s_{\lambda}^{2}}\right]
\end{equation}\label{Eq:test}

In the following section, we explain the quality cuts used to choose a high quality training set in order to generate the required \textit{Cannon} spectral model as described above.

\begin{figure*}
	\includegraphics[width=\textwidth]{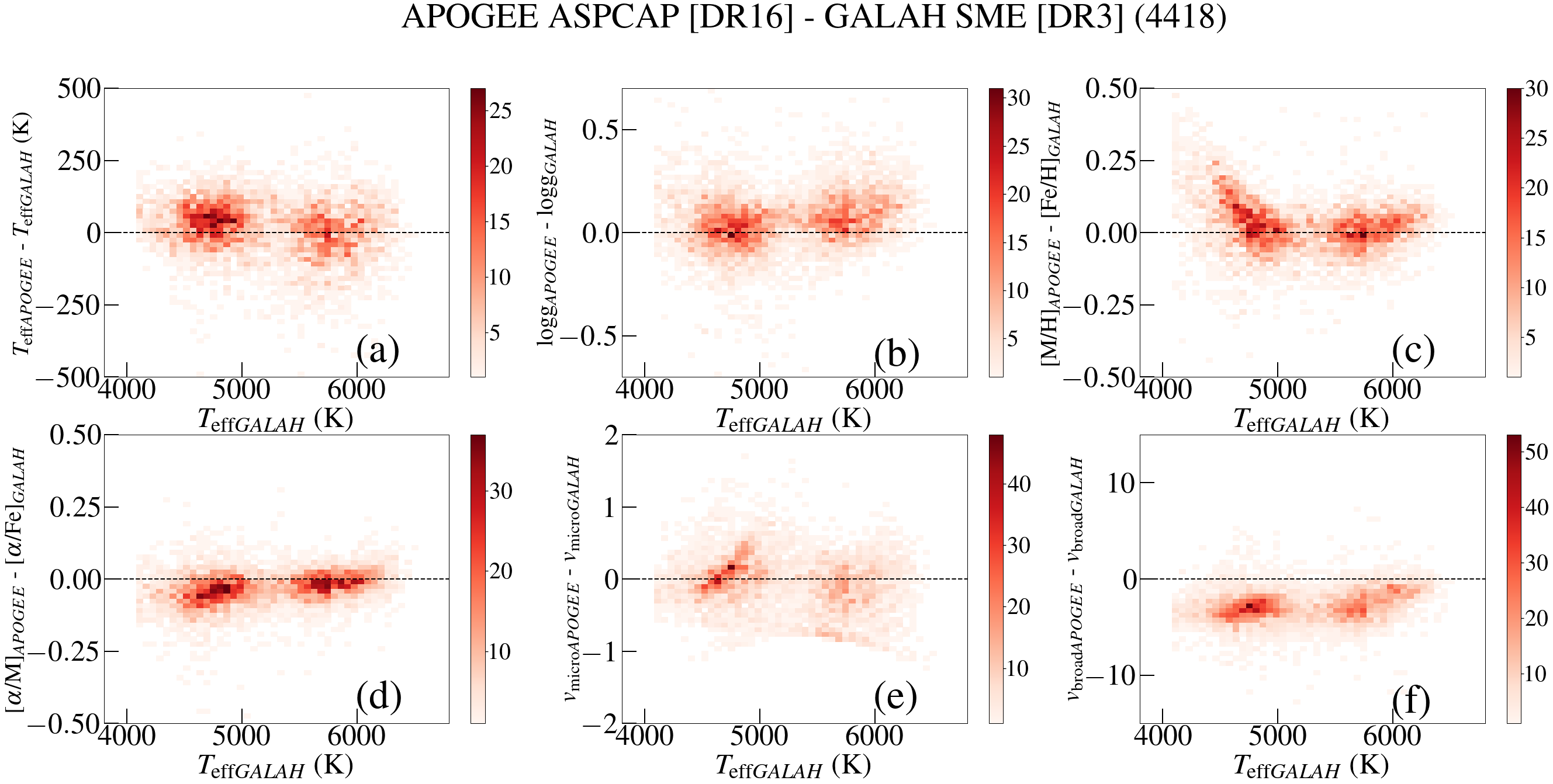}
    \caption{The difference in derived stellar parameters as a function of T$_\mathrm{eff}$ from GALAH SME values for 4418 stars (shown as 2d histogram colour coded by the number of stars in each bin) in the training set. The two clumps, cool (4000 $<$ T$_\mathrm{eff}$ $<$ 5200 K) and hot (5200 $<$ T$_\mathrm{eff}$ $<$ 6200 K), seen for all parameters approximately represent difference trend for giants and dwarfs respectively}. All six parameters used as input labels in the \textit{Cannon} are shown: (a) T$_\mathrm{eff}$, (b) $\log g$, (c) [Fe/H], (d) [$\alpha$/Fe], (e) $v_{\rm micro}$ and (f) $v_{\rm broad}$.      
    \label{fig:Sysdiff_trset_ApgGalah}
\end{figure*}

\begin{table*}
\caption{Mean and scatter values of APOGEE-GALAH difference for six stellar labels in the training set. }\label{table:sys}
\begin{tabular}{c c c c c c c }
\hline
\hline
 APOGEE - GALAH & \multicolumn{2}{c}{All} & \multicolumn{2}{c}{Giants}  & \multicolumn{2}{c}{Dwarfs}   \\
 \hline
  Parameter (unit) & $\mu$ & $\sigma$   & $\mu$ & $\sigma$ & $\mu$ & $\sigma$     \\
\hline
\centering{T$_\mathrm{eff}$} (K) & 14 & 84 & 35 & 68 & 0 & 94 \\
$\log g$ (dex) & 0.06 & 0.12 & 0.03 & 0.13 & 0.08 & 0.10 \\
 $[Fe/H]$ (dex) & 0.02 & 0.08 & 0.05 & 0.10 &  0.00 & 0.06 \\
 $[\alpha/Fe]$ (dex) & -0.03 & 0.05 & -0.04 & 0.06 & -0.02 & 0.05 \\
 $v_{\rm micro}$ (km/s) & -0.05 & 0.37 & 0.07 & 0.30 & -0.14 & 0.40 \\
 $v_{\rm broad}$ (km/s) & -2.80 & 1.50 & -3.08 & -1.21 & -2.59 & 1.62 \\
\hline
\hline
\end{tabular}
\end{table*}

\subsection{Training set}
\label{sec:trset}

We construct one training set that can be used to carry out 4 combinations of cross-survey \textit{Cannon} training and modeling. Hence, we use the GALAH DR3 combined spectra catalogue with one entry per star for which spectra from multiple exposures have been stacked to estimate their stellar parameters and elemental abundances. In the case of APOGEE, we choose stars with stellar parameters derived from high SNR  combined spectrum. Similarly, we choose combined spectra from both surveys\footnote{For APOGEE, combined spectra is provided as the HDU1 extension of 'apStar' fits files. For GALAH, combined spectra fits files are named after their 'sobject$\_$id' from which the pseudo continuum normalized flux can be extracted.} for constructing the training set and later in testing stage where predictions are made.

 We find 14,406 stars based on cross match using the \textit{APOGEE$\_$ID} and \textit{star$\_$id} columns in APOGEE DR16 and GALAH DR3 catalogues with \textit{Topcat} (\citealt{Taylor:2005,Taylor:2020}). We remove stars with invalid GALAH and APOGEE stellar parameters and further select reliable and high quality survey labels/stellar parameters by certain constraints for each survey.
 
For GALAH, we choose stars with SNR ratio of spectra in the green arm (\textit{snr$\_$c2$\_$iraf}) $>$ 25, chi-square value of stellar parameter fitting the following constraints (\textit{chi2$\_$sp}) $<$ 4 and the flag that describes various GALAH reduction and analysis issues indicating the quality of spectra and estimated stellar parameters (\textit{flag$\_$sp}) is equal to zero.

For APOGEE, we choose stars with SNR $>$ 80 in addition to removing stars for which selected bits (16: bad T$_\mathrm{eff}$, 17: bad $\log g$, 18: bad $v_{\rm micro}$, 19: bad metals, 20: bad [$\alpha$/Fe] and 23: bad overall for star) in the \textit{ASPCAPFLAG}\footnote{https://www.sdss.org/dr14/algorithms/bitmasks/} has been set.

As for selecting good quality spectra, we neglect spectra with \textit{STARFLAG}\footnote{https://www.sdss.org/dr14/algorithms/bitmasks/$\#$APOGEE\_STARFLAG} bits set for selected few bits (0: bad pixels, 3: very bright neighbor, 4: low snr, 9: significant number of pixels in high persistence region, 10: significant number of pixels in medium persistence region , 11: significant number of pixels in low persistence region, 12: obvious positive jump in blue chip, 13: obvious negative jump in blue chip and 17: Broad lines). 

Figure~\ref{fig:HR_alphafe} shows the Kiel diagrams for all stars in GALAH (left panel) and APOGEE (right panel) overlaid with the training set contours with respective pipeline estimates in black. In Fig~\ref{fig:Sysdiff_trset_ApgGalah}, we show the systematic difference trends in T$_\mathrm{eff}$, $\log g$, [Fe/H], [$\alpha$/Fe], $v_{\rm micro}$ and $v_{\rm broad}$ for these stars, with T$_\mathrm{eff}$$_{\rm, GALAH}$ on the x-axis and APOGEE-GALAH values on the y-axis. The two clumps, cool (4000 $<$ T$_\mathrm{eff}$ $<$ 5200 K) and hot (5200 $<$ T$_\mathrm{eff}$ $<$ 6200 K), seen for all parameters approximately represent difference trend for giants and dwarfs respectively. Table~\ref{table:sys} lists the mean and standard deviation (calculated as the mid value of 84$^{th}$-16$^{th}$ percentile to get rid of outliers) values for all stars, giants and dwarfs. In this case giants and dwarfs are selected based on $\log g$ cut of 3.5.



Among the six stellar parameters, we find significant differences/offsets in the case of [Fe/H], $v_{\rm micro}$ and $v_{\rm broad}$. The differences for all parameters have a non zero mean value and exhibits various trends with T$_\mathrm{eff}$. APOGEE temperatures are higher and derives higher $\log g$ values for hot dwarfs compared to GALAH. Meanwhile cool giants in APOGEE have lower [$\alpha$/Fe] measurements compared to GALAH.

The giant and dwarf clumps in the Figure~\ref{fig:Sysdiff_trset_ApgGalah}c hints at different systematic trends of [Fe/H] with respect to T$_\mathrm{eff}$ for giants and dwarfs. There is a steep declining trend for giants, with higher metallicity values ( $\sim$0.4 dex relative to GALAH abundances) measured by APOGEE for cool giants (T$_\mathrm{eff}$ $<$ 4500 K). This trend is similar to one of the possible caveats mentioned in GALAH DR3 (Section 6.5 of \cite{GALAHDR3}). They have noticed a significant trend of underestimated [Fe/H] with increasing metallicity, when comparing with GALAH DR2, for the metal-rich ([Fe/H $>$ 0) giants and red clump stars. As discussed in \cite{GALAHDR3}, the reasons for this could be many fold, e.g, missing/unreliable molecular line data, the underestimation of blending and incorrect continuum normalisation, over/under estimation of $v_{\rm micro}$ etc. For dwarfs, there is better agreement between APOGEE and GALAH metallicities with a slight increase (upto $\sim$ 0.15-0.2 dex) in APOGEE metallicity for hotter stars (T$_\mathrm{eff}$ $>$ 6000 K).
\\
In Figure~\ref{fig:Sysdiff_trset_ApgGalah}e, we find significant scatter ( $>$ $\sim$ 0.30 km/s) in the $v_{\rm micro}$ difference between the two surveys, with mean differences of -0.05 km/s, 0.07 km/s and -0.14 km/s for the whole sample, giants and dwarfs respectively. The difference in trends for giants and dwarfs are also evident from the two clumps. Such a significant difference may be attributed to the way in which $v_{\rm micro}$ is determined in GALAH and APOGEE. While empirical relations (Equations 4 and 5 in \citealt{GALAHDR3}) are employed in the case of GALAH, APOGEE uses $v_{\rm micro}$ as a free parameter while determining synthetic spectra that best fits the observed spectra.

In Figure~\ref{fig:Sysdiff_trset_ApgGalah}f, a systematic difference in $v_{\rm broad}$ can be seen with the APOGEE values consistently lower than GALAH values. This may again be attributed to the difference in $v_{\rm broad}$ determination in both surveys. In GALAH, $v_{\rm broad}$ is determined using SME by setting $v_{\rm mac}$ (macroturbulent velocity) to 0 and only fitting for $v \sin i$ (rotational velocity). In APOGEE, $v_{\rm broad}$ for dwarfs are the $v \sin i$ values estimated in the same way as in the case of $v_{\rm micro}$ using a grid with 7 steps of $v \sin i$ (1.5, 3.0, 6.0, 12.0, 24.0, 48.0, 96.0 km/s), while for giants an empirical relation is used to estimate $v_{\rm mac}$ \citep{Jonsson:2020}.

Even though there is reasonably good agreement between the two surveys (especially for T$_\mathrm{eff}$ and $\log g$) in Figure~\ref{fig:Sysdiff_trset_ApgGalah}, there are systematic differences in [Fe/H], [$\alpha$/Fe], $v_{\rm micro}$ and $v_{\rm broad}$ for the same stars in both surveys. As mentioned in Section~\ref{sec:data}, both surveys observe stars in different wavelength regimes (optical for GALAH, near infrared for APOGEE) and employ different methodology and pipelines to estimate these parameters which could result in such systematic differences. These systematic differences for same stars as well as contrasting difference trends for giants and dwarfs thus show the importance of placing these surveys on the same abundance scales if they are to be used in conjunction. This also emphasises the need for observation of larger number of common stars between large scale spectroscopic surveys, which will enable cross survey calibrations as well as more consistent analysis pipelines in the future.

\begin{figure*}
	\includegraphics[width=\textwidth]{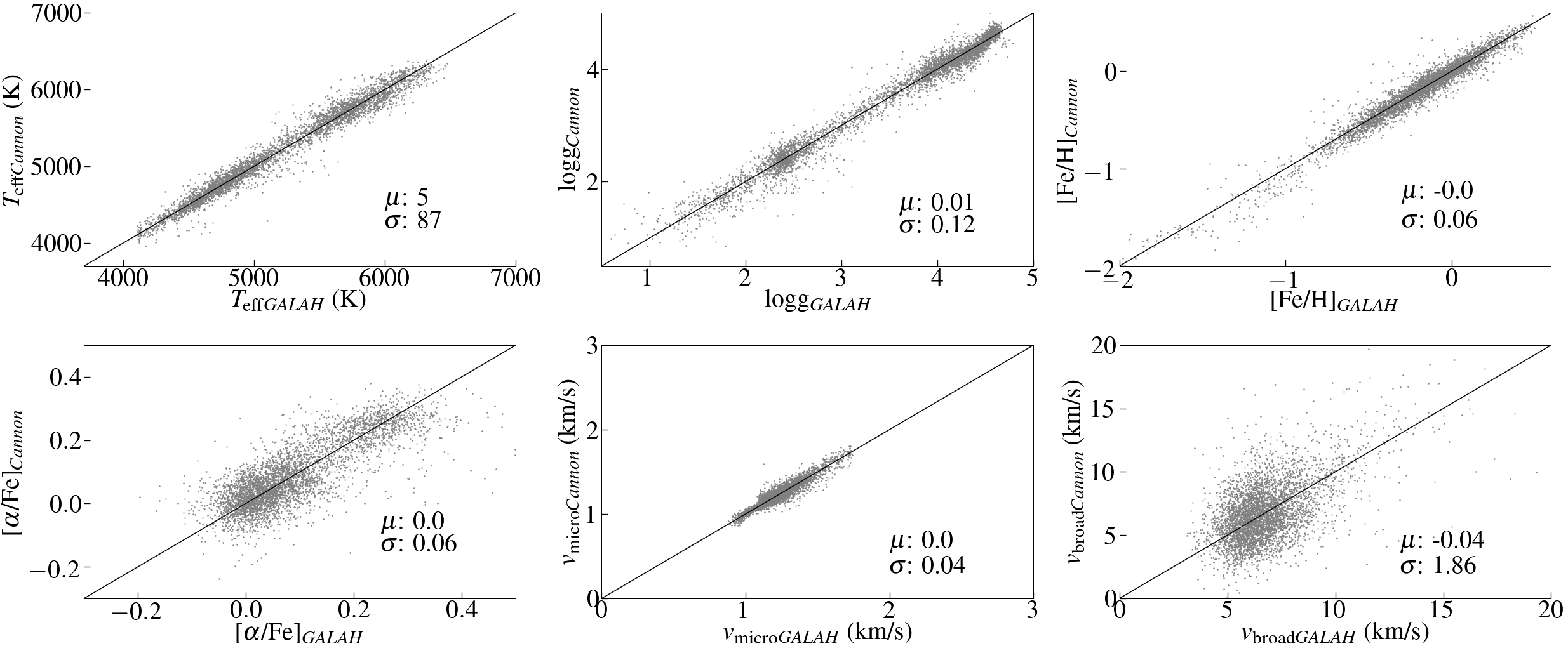}
    \caption{One-to-one relation of GALAH SME stellar labels (x-axis) in the training set versus corresponding values estimated by the \textit{Cannon} from APOGEE spectra (y-axis) after carrying out a 12-fold cross-validation test. The mean and scatter (calculated as the mid value of 84$^{th}$-16$^{th}$ percentile) in each label difference is indicated on the bottom right hand side of each plot. }
    \label{fig:CV_ApgGalah}
\end{figure*}

\begin{figure*}
	\includegraphics[width=\textwidth]{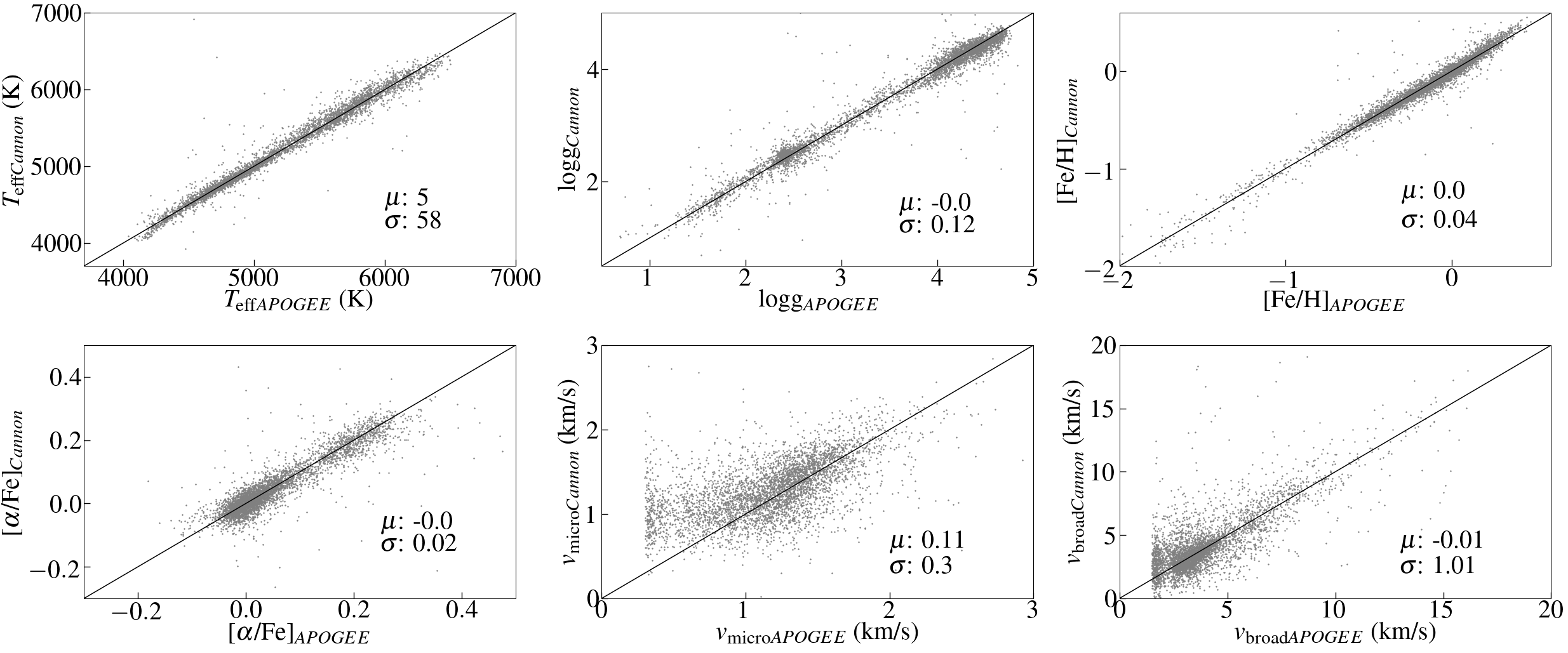}
    \caption{One-to-one relation of APOGEE ASPCAP stellar labels (x-axis) in the training set versus corresponding values estimated by the \textit{Cannon} from GALAH spectra (y-axis) after carrying out a 12-fold cross-validation test. The mean and scatter (calculated as the mid value of 84$^{th}$-16$^{th}$ percentile) in each label difference is indicated on the bottom right hand side of each plot.  }
    \label{fig:CV_GalahApg}
\end{figure*}

\begin{figure*}
	\includegraphics[width=\textwidth]{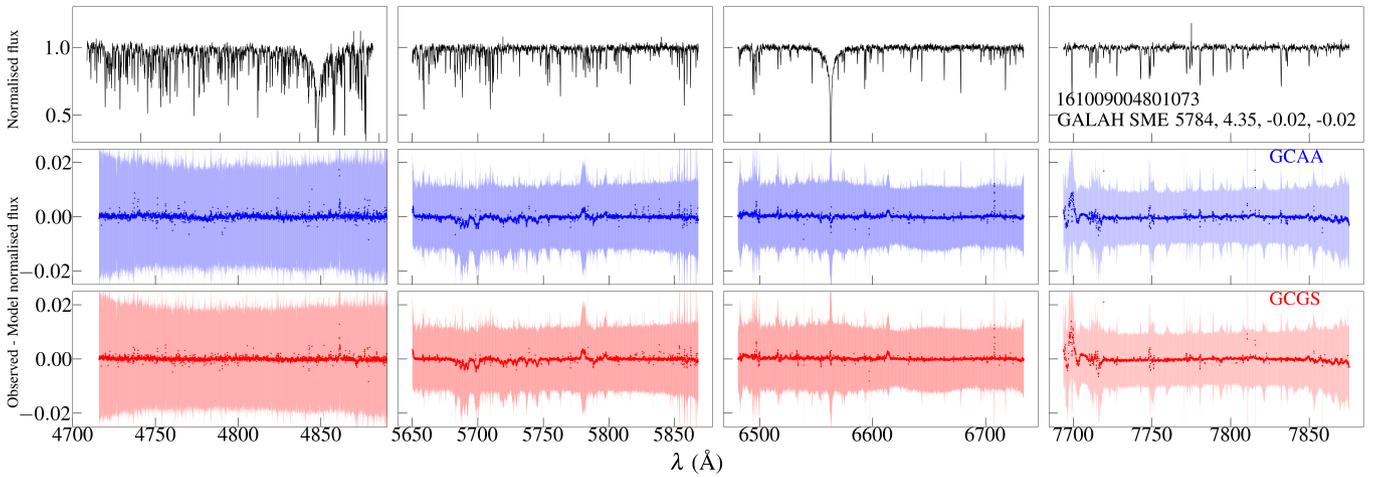}
    \caption{Pixel-by-pixel comparison of the \textit{Cannon} model (GCAA and GCGS) spectra with the observed GALAH spectra for all 4,418 stars in the training set. The top row shows a typical solar-type GALAH star spectrum with the GALAH SME stellar parameters (T$_\mathrm{eff}$, $\log g$, [Fe/H] and [$\alpha$/Fe]) listed in the right most panel. The middle and bottom rows shows the comparison for GCAA and GCGS respectively with the full GALAH wavelength ranges in the 4 chips plotted in 4 panels from left to right. The solid scatter points represent the median residual value ( observed - model normalised flux) estimated at each pixel and the band around each point/pixel denote the 16$^{th}$ and 84$^{th}$ percentile values of residuals estimated at the respective pixel. Median residuals are very close to 0 for majority of pixels, while higher residuals are seen at either bad/noisy pixels or pixels that correspond to lines of elements (e.g. Cr at $\sim$5700 \AA, Cu at $\sim$5782 \AA, and strong K line at $\sim$7700 \AA) that haven't been modelled in our \textit{Cannon} models.     }
    \label{fig:spectra_GS}
\end{figure*}

\begin{figure*}
	\includegraphics[width=\textwidth]{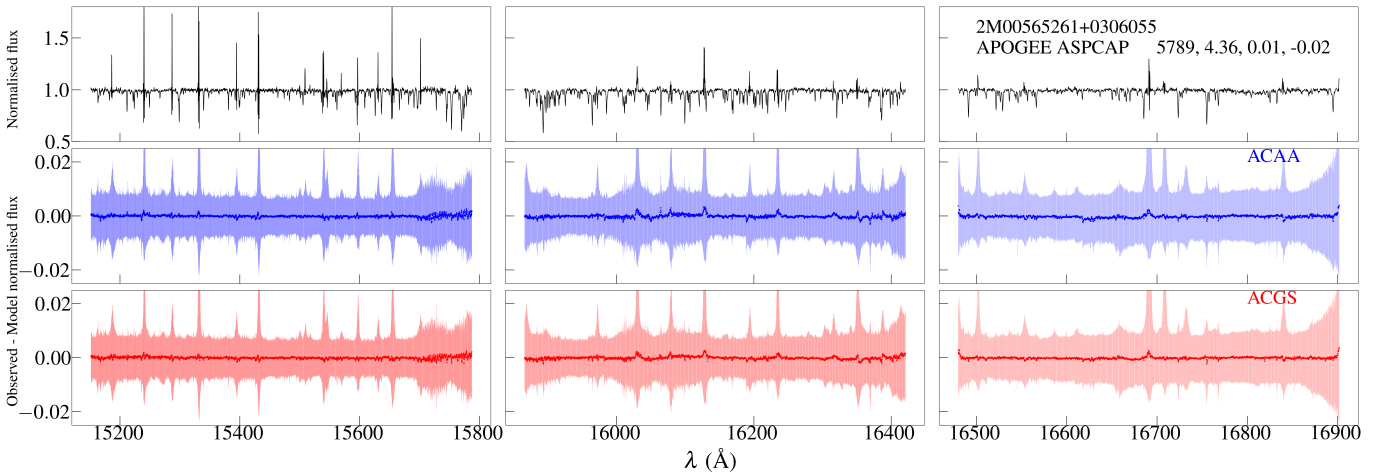}
    \caption{ Pixel-by-pixel comparison of the \textit{Cannon} model (ACAA and ACGS) spectra with the observed APOGEE spectra for all 4,418 stars in the training set. The top row shows a typical solar-type APOGEE star spectrum with the APOGEE ASPCAP stellar parameters (T$_\mathrm{eff}$, $\log g$, [Fe/H] and [$\alpha$/Fe]) listed in the right most panel. The middle and bottom rows shows the comparison for ACAA and ACGS respectively with the full GALAH wavelength ranges in the 3 wavelength bands plotted in 3 panels from left to right. The solid scatter points represent the median residual value ( observed - model normalised flux) estimated at each pixel and the band around each point/pixel denote the 16$^{th}$ and 84$^{th}$ percentile values of residuals estimated at the respective pixel. As in the case of GALAH, median residuals are very close to 0 for majority of pixels. Higher residuals are dominated by bad/noisy pixels which is evident on comparison with the solar-type star spectrum in the top panels.}     
    \label{fig:spectra_AA}
\end{figure*}

\begin{table*}
\caption{Naming convention used in this work to indicate 4 catalogues resulting from the use of two surveys and their spectra in different combinations.  }\label{table:naming}
\begin{tabular}{c c c }
\hline
\hline
 Name & Spectra & Labels  \\
\hline
\textit{APOGEE Cannon GALAH SME (ACGS)} & APOGEE & GALAH SME \\
\textit{APOGEE Cannon APOGEE ASPCAP (ACAA)} & APOGEE & APOGEE ASPCAP \\
\textit{GALAH Cannon APOGEE ASPCAP (GCAA)} & GALAH & APOGEE ASPCAP \\
\textit{GALAH Cannon GALAH SME (GCGS)} & GALAH & GALAH SME \\
\hline
\hline
\end{tabular}
\end{table*}


\subsection{Training }
\label{sec:Train}

Once the training set is finalised, we proceed to carry out the training and cross-validation. As mentioned in Section \ref{sec:intro}, the objective of this work is to provide two combined stellar parameter catalogs of APOGEE and GALAH, one scaled in terms of APOGEE and the other in terms of GALAH. Hence we use spectra and labels from both surveys in 4 different combinations, starting from the training step. To avoid any confusion resulting from this, hereafter we introduce a naming convention to indicate each case in Table~\ref{table:naming}:

In the following section, we focus on APOGEE Cannon GALAH SME (ACGS) and GALAH Cannon APOGEE ASPCAP (GCAA), while similar exercises for APOGEE Cannon APOGEE ASPCAP (ACAA) and GALAH Cannon GALAH SME (GCGS) are explained in the Appendix~\ref{app:Galah}.


We limit the labels that we train and infer to T$_\mathrm{eff}$, $\log g$, [Fe/H], [$\alpha$/Fe], $v_{\rm micro}$, and $v_{\rm broad}$ where [Fe/H] and [$\alpha$/Fe] refer to the general metallicity and alpha abundance labels in each survey. We do not go beyond the general alpha abundance to include individual elements as in such cases, we need to know all abundances for each training set entry - which becomes an increasingly difficult task with an increasing number of elements. In particular, elements other than the alpha-elements, like Li or s- and r-process elements are impossible to measure always throughout the whole parameter space. Various challenges involved in determination of more elements using the \textit{Cannon} is discussed in detail in \cite{Buder:2018}.


While we use the whole spectral wavelength range for training and determining label coefficients for T$_\mathrm{eff}$, $\log g$, [Fe/H], $v_{\rm micro}$, and $v_{\rm broad}$, we make use of censoring/line masks in the case of [$\alpha$/Fe]. This is similar to the use of line masks in SME to estimate abundances for each element in a spectrum. [$\alpha$/Fe] for GALAH is determined from an error-weighted combination of selected lines of Si, Mg, Ti and Ca, while APOGEE [$\alpha$/M] is determined from a combination of O, Mg, Si, S, Ca and Ti lines. We decided to use line masks for the common elements in both surveys (Si, Mg, Ti and Ca) to be used as censors in the training step. This ensures that \textit{Cannon} does not incorporate incorrect correlations from other lines in the respective spectra while predicting [$\alpha$/Fe] values. In the case of GALAH, the line masks for Si, Mg, Ti and Ca are available in the linelist used to determine respective abundances with SME ( Table. A2 in \citealt{GALAHDR3}). For APOGEE, \cite{Jonsson:2020} provides the windows and weights used in the determination of stellar abundances in their Table 3. We select the wavelength windows for Si, Mg, Ti, Ca lines. 

In the training step, 4 \textit{Cannon} models corresponding to ACGS, GCGS, GCAA and ACAA are created using spectra and labels of stars in the training set. Coefficient matrix, $\theta_{\lambda}$, and scatter,$s_{\lambda}$, are obtained by optimising Equation~\ref{Eq:train}, given the labels in each case.

\begin{figure*}
	\includegraphics[width=\textwidth]{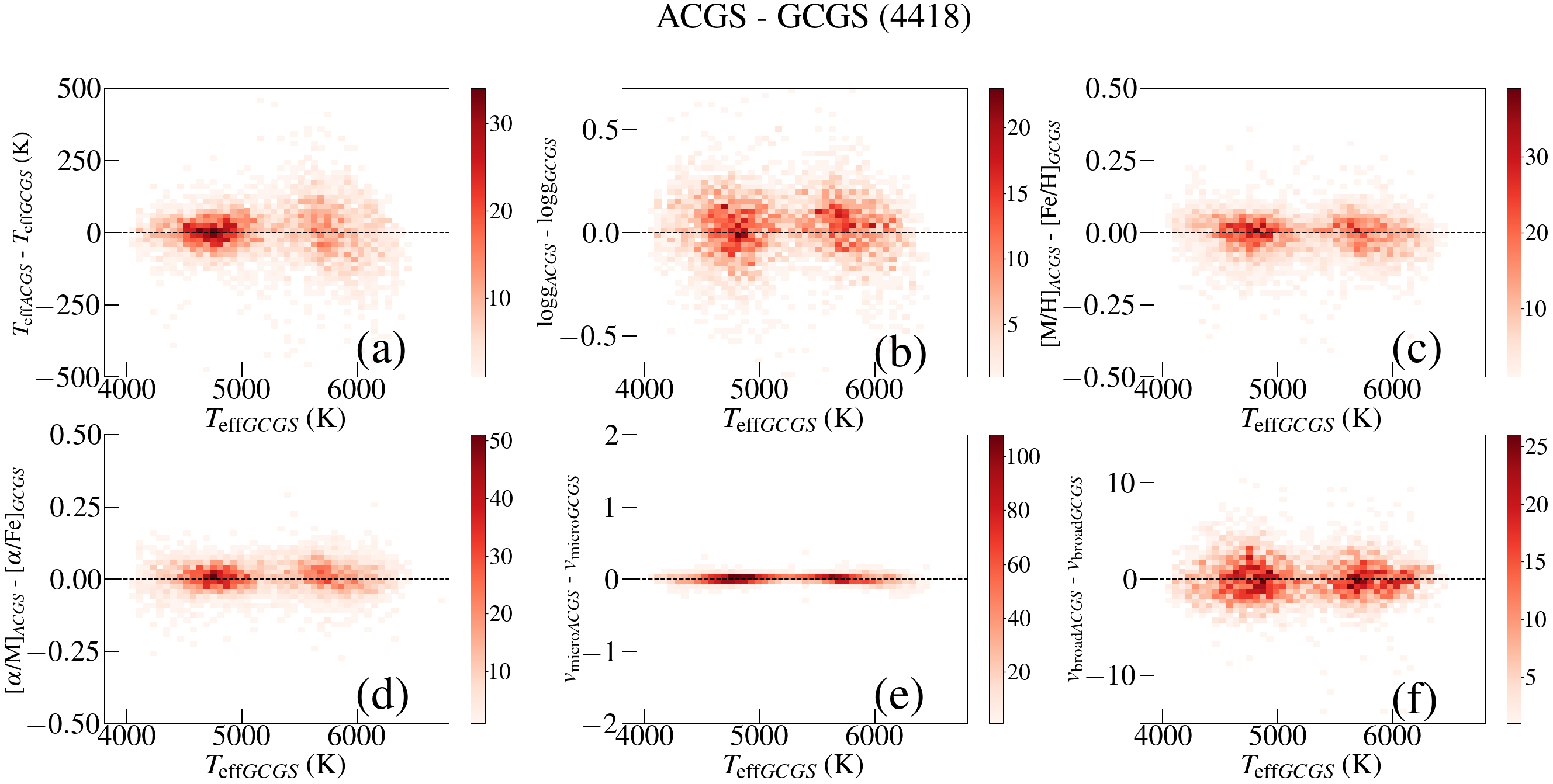}
    \caption{ Same as Figure~\ref{fig:Sysdiff_trset_ApgGalah} except for label differences between training set stars in ACGS and GCGS. Compared to differences seen in Figure~\ref{fig:Sysdiff_trset_ApgGalah}, metallicity and alpha abundance differences have decreased scatter and less trends, demonstrating the effectiveness of the \textit{Cannon} models.} 
    \label{fig:sysdiff_ACGS_GCGS}
\end{figure*}

\begin{figure*}
	\includegraphics[width=\textwidth]{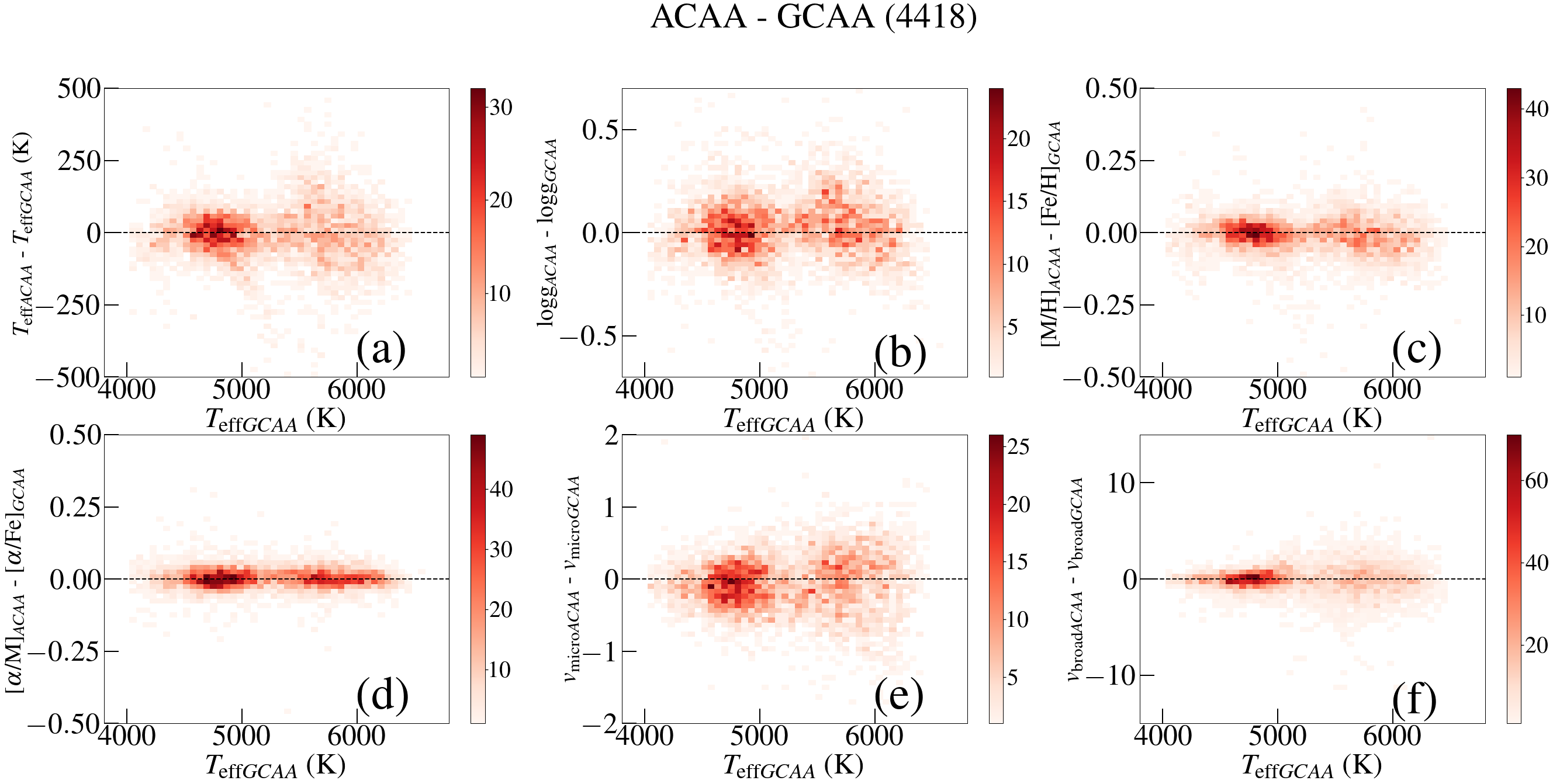}
    \caption{ Same as Figure~\ref{fig:Sysdiff_trset_ApgGalah} except for label differences between training set stars in ACAA and GCAA. Reduced scatter and not less trends are seen here, with the mean difference always lying close to 0. This demonstrates the effectiveness of our method and the \textit{Cannon} models we used.}
    \label{fig:sysdiff_ACAA_GCAA}
\end{figure*}

\subsection{Cross-Validation}
\label{sec:Cross-Validation}

In order to verify how well the\textit{ Cannon} model is able to recover the training set labels, we carry out a 12-fold cross-validation. For this, we create twelve random subgroups from the training set, then use them individually as test set while the remaining eleven are treated as the training data. We show the resulting one-to-one relation, i.e., predicted label values vs input label values for ACGS and GCAA in Figures \ref{fig:CV_ApgGalah} and \ref{fig:CV_GalahApg} respectively. 

When we train the \textit{Cannon} model on APOGEE spectra using GALAH SME labels (ACGS; Figure~\ref{fig:CV_ApgGalah}), all labels except $v_{\rm broad}$ and [$\alpha$/Fe] follow a tight one-to-one relation. T$_\mathrm{eff}$ relation is tight with mean difference of $\sim$5 K and scatter of $\sim$ 90 K. The \textit{Cannon} estimates for $\log g$ and [Fe/H] values are similar to corresponding input GALAH labels, while larger dispersion in input GALAH [$\alpha$/Fe] values is reflected in the \textit{Cannon} output as well. The tightest one-to-one relation is seen in the case of $v_{\rm micro}$, for which the input GALAH labels are estimated using empirical relations. As this relation is a function of GALAH T$_\mathrm{eff}$ and $\log g$, these values are well correlated with respective spectra in the training step and thus ensures that the \textit{Cannon} model is able to reproduce them from APOGEE spectra in the label inference step. The largest scatter is seen in the case of $v_{\rm broad}$, meaning that the \textit{Cannon} model is not able to correlate the input GALAH $v_{\rm broad}$ values with the features in APOGEE spectra. At the same time, we find that the \textit{Cannon} model trained on GALAH spectra using GALAH labels (GCGS) are able to estimate similar $v_{\rm broad}$ values from GALAH spectra in the label inference step with less scatter (see Appendix, Figure~\ref{fig:CV_ApgGalah_GALAH}). Thus the reason for the large dispersion seen here in the case of $v_{\rm broad}$ is not clear.

When we train the \textit{Cannon} model on GALAH spectra using APOGEE ASPCAP labels (Figure~\ref{fig:CV_GalahApg}), the \textit{Cannon} output values for T$_\mathrm{eff}$, $\log g$, [Fe/H] and [$\alpha$/Fe] is as seen in ACGS, with tight one-to-one relations with the respective APOGEE ASPCAP values. There are significant deviations in the case of $v_{\rm micro}$ and $v_{\rm broad}$. The \textit{Cannon} outputs for input $v_{\rm micro}$$_{\rm APOGEE}$ $>$ 1 km/s follow one-to-one relation, whereas the \textit{Cannon} estimates are higher for lower $v_{\rm micro}$$_{\rm APOGEE}$ values. This indicates that the \textit{Cannon} model is unable to find significant correlation between $v_{\rm micro}$$_{\rm APOGEE}$ values and corresponding GALAH spectra in this lower $v_{\rm micro}$$_{\rm APOGEE}$ range. Much tighter one-to-one trend is seen in the case of $v_{\rm broad}$, though there are significant deviations for lower $v_{\rm broad}$$_{\rm APOGEE}$ values. When we use the \textit{Cannon} model trained on APOGEE spectra using APOGEE labels (ACAA), the resulting \textit{Cannon} $v_{\rm micro}$ and $v_{\rm broad}$ estimates also show significant deviations in similar $v_{\rm micro}$ and $v_{\rm broad}$ ranges (see Appendix~\ref{app:CV}). This may be because of the limited sensitivity at such low values of $v_{\rm micro}$ and $v_{\rm broad}$.

In all cases and for all labels, there are deviations from the one-to-one line close to the high and low values of the respective labels, i.e., the label space edges. This most likely shows the inability of the \textit{Cannon} to interpolate at the training set edges.

Overall, the 12-fold cross-validation shows the capability of the \textit{Cannon} models trained on APOGEE (GALAH) spectra with GALAH (APOGEE) labels (with training set comprising good quality stars commonly observed by both surveys) to infer GALAH (APOGEE) scaled values of T$_\mathrm{eff}$, $\log g$, [Fe/H] and [$\alpha$/Fe] from APOGEE (GALAH) spectra.

\subsection{Application to GALAH and APOGEE}
\label{sec:test}
We proceed to use the\textit{ Cannon} models after training and cross-validation to predict the GALAH scaled labels for 437,445 APOGEE spectra and APOGEE scaled labels for 576,390 GALAH spectra of unique stars. Using 4 \textit{Cannon} spectral models corresponding to ACGS, GCGS, GCAA and ACAA, all six labels are estimated for each case by optimising the log likelihood function in Equation~\ref{Eq:test}, given the respective coefficient matrix, $\theta_{\lambda}$, and scatter,$s_{\lambda}$, from the training step. 

In the following sub sections, we demonstrate the effectiveness of our \textit{Cannon} models by comparing model spectra with observed spectra (both GALAH and APOGEE) and plotting the systematic differences of all labels for common stars in APOGEE and GALAH.

\begin{table*}
\caption{Mean and scatter values in ACGS - GCGS and ACAA - GCAA difference of six stellar labels for common stars in APOGEE and GALAH.} \label{table:sys_cannon}
\begin{tabular}{c c c c c }
\hline
\hline
  & \multicolumn{2}{c}{ACGS - GCGS} & \multicolumn{2}{c}{ACAA - GCAA}  \\
 \hline
  Parameter (unit) & $\mu$ & $\sigma$   & $\mu$ & $\sigma$      \\
\hline
\centering{T$_\mathrm{eff}$} (K) & 2 & 66 & -4 & 73 \\
$\log g$ (dex) & 0.03 & 0.12 & 0.02 & 0.12 \\
 $[Fe/H]$ (dex) & 0.00 & 0.05 & 0.00 & 0.05 \\
 $[\alpha/Fe]$ (dex) & 0.01 & 0.04 & 0.00 & 0.02  \\
 $v_{\rm micro}$ (km/s) & 0.00 & 0.03 & -0.06 & 0.34  \\
 $v_{\rm broad}$ (km/s) & -0.01 & 1.78 & -0.06 & 1.07 \\
\hline
\hline
\end{tabular}
\end{table*}

\begin{table*}
\caption{Scaling factor for each label covariance error estimated from repeat observations using Equation~\ref{eq:err} for each catalogue.  }\label{table:ErrRepeat}
\begin{tabular}{c c c c c}
\hline
\hline
 Parameter (unit) & ACGS & GCGS & GCAA & ACAA   \\
\hline 
\centering{T$_\mathrm{eff}$} (K) & 12200 & 7200 & 6925 & 11325\\
$\rm \log g$ (dex) & 18 & 8.6 & 8.5 & 19 \\
 $\rm [Fe/H]$ (dex) & 8.9 & 6.4 & 7.7 & 9 \\
 $\rm [\alpha/Fe]$ (dex) & 1.2 & 0.5 & 0.4 & 1.2 \\
 $v_{\rm micro}$ (km/s) & 3.8 & 2.1 & 5.3 & 12 \\
 $v_{\rm broad}$ (km/s) & 52.1 & 31.5 & 29 & 56 \\
\hline
\hline
\end{tabular}
\end{table*}

\subsubsection{ \textit{Cannon} model spectra}
\label{sec:spectra}

Here we compare observed survey spectra with the spectra generated by the \textit{Cannon} models to demonstrate how well \textit{Cannon} spectral models reproduce the observed spectra. For this we first estimate the residual values (observed - model normalised flux) for all 4,418 stars in the training set at each wavelength/pixel. We then plot the median residual value along with the respective 16$^{th}$ and 84$^{th}$ percentile values at each pixel as shown in Figures~\ref{fig:spectra_GS} and ~\ref{fig:spectra_AA} for model spectra comparisons with GALAH spectra and APOGEE spectra respectively. In both the figures, we show observed solar-type star spectrum from the respective survey in the top row to help the readers identify lines and the residual plots for APOGEE scaled and GALAH scaled cases respectively in the top and bottom rows with the full survey wavelength ranges plotted in the panels from left to right. The solid scatter points represent the median residual value estimated at each pixel and the band around each point/pixel denote the 16$^{th}$ and 84$^{th}$ percentile values of residuals at the respective pixel.

Median values for majority of pixels are very close to 0 with the 16$^{th}$ and 84$^{th}$ percentile values lying within 0.01, except in the case of GALAH chip 1 which is found to be inherently noisy. In the case of GALAH (Figure~\ref{fig:spectra_GS}), slightly higher residual values are seen at seen at either bad/noisy pixels or pixels that correspond to lines of elements (e.g. Cr at $\sim$5700 \AA, Cu at $\sim$5782 \AA, and strong K line at $\sim$7700 \AA) that have not been modelled in our \textit{Cannon} models. Comparing the pixel positions of higher residual values with lines in APOGEE solar-type star spectrum, it is evident that the bad/noisy spike pixels are the dominant reason for higher residuals in the case of APOGEE (Figure~\ref{fig:spectra_AA}). 

This demonstrates the ability of the \textit{Cannon} spectral models to reproduce the observed spectra for stars in the parameter space covered by our training set .


\subsubsection{ Systematic difference using common stars}
\label{sec:commonstars}

In order to demonstrate the effectiveness of our method, we compare GALAH scaled labels from APOGEE spectra (ACGS) and GALAH spectra (GCGS) for common stars which are in the training set. For the same stars, we compare APOGEE scaled labels from APOGEE spectra (ACAA) and GALAH spectra (GCAA). This is shown in Figures~\ref{fig:sysdiff_ACGS_GCGS} and ~\ref{fig:sysdiff_ACAA_GCAA} where we plot the differences in all six labels as a function of T$_\mathrm{eff}$ for ACGS - GCGS and ACAA - GCAA respectively. This is similar to the Figure~\ref{fig:Sysdiff_trset_ApgGalah} where we showed the differences between the pipeline values from APOGEE and GALAH. Table~\ref{table:sys_cannon} lists the mean and scatter of the difference values for all labels in each case. On comparison with the mean and scatter values for APOGEE-GALAH listed in Table~\ref{table:sys}, we find reduced scatter and less trends for common stars in both surveys from GALAH scaled and APOGEE scaled catalogues. The improvement is especially evident in metallicity and alpha abundance with the mean difference close to 0 and the absence of any significant trends.

\subsection{Flagging}
\label{sec:flag}

As a first step, we need to flag \textit{Cannon} estimates that lie outside the bounds of the training set labels since the \textit{Cannon} cannot reliably extrapolate to different regimes outside of the training set. For each test set spectra, we estimate the distance, D, of the test set \textit{Cannon} estimate, $\rm l_{test}$, to the training set labels, $l_{nTs}$, in similar way as described in \cite{Buder:2018} and \cite{Ho:2017} :

\begin{equation}
    D = \sum_{l}\sum_{nTs}\frac{(l_{test} - l_{nTs})}{K_{l}^{2}}
\end{equation}

\noindent where $K_{l}$ represent the uncertainties in each label for which we use RMS values obtained from cross-validation step in Section~\ref{sec:Cross-Validation}. Using T$_\mathrm{eff}$, $\log g$, [Fe/H], [$\alpha$/Fe] as the label space, \textit{l}, we calculate the average distance to the closest 10 stars in the training set and flag those \textit{Cannon} estimates that are farther than 8 (2$\sigma$ for 4 labels). We indicate this in the final catalogue by setting the \textit{flag$\_$Cannon$\_$dist} to 1.  
In addition, we provide \textit{flag$\_$spectra} to indicate problematic spectra from APOGEE (STARFLAG bits as explained in Section\ref{sec:trset}) and GALAH, \textit{flag$\_$sp$\_$aspcap} to indicate stars that have been flagged by respective surveys (flag$\_$sp in GALAH and ASPCAP FLAG in APOGEE) and \textit{flag$\_$survey} to indicate the stars with invalid values (from ASPCAP and SME) for the labels we use. \textit{flag$\_$sp$\_$aspcap} follows the same format as in the respective survey. \textit{flag$\_$spectra} for APOGEE stars follow the format in APOGEE DR16 catalogue, while value of 1 is used to indicate bad GALAH spectra. With these flags, we indicate those stars/spectra that has corresponding issues and emphasize that it is better to avoid these stars for exploring the Milky Way chemo-dynamic trends. Table~\ref{table:scheme} lists and describe all the above mentioned flags.

\begin{figure*}
	\includegraphics[width=\textwidth]{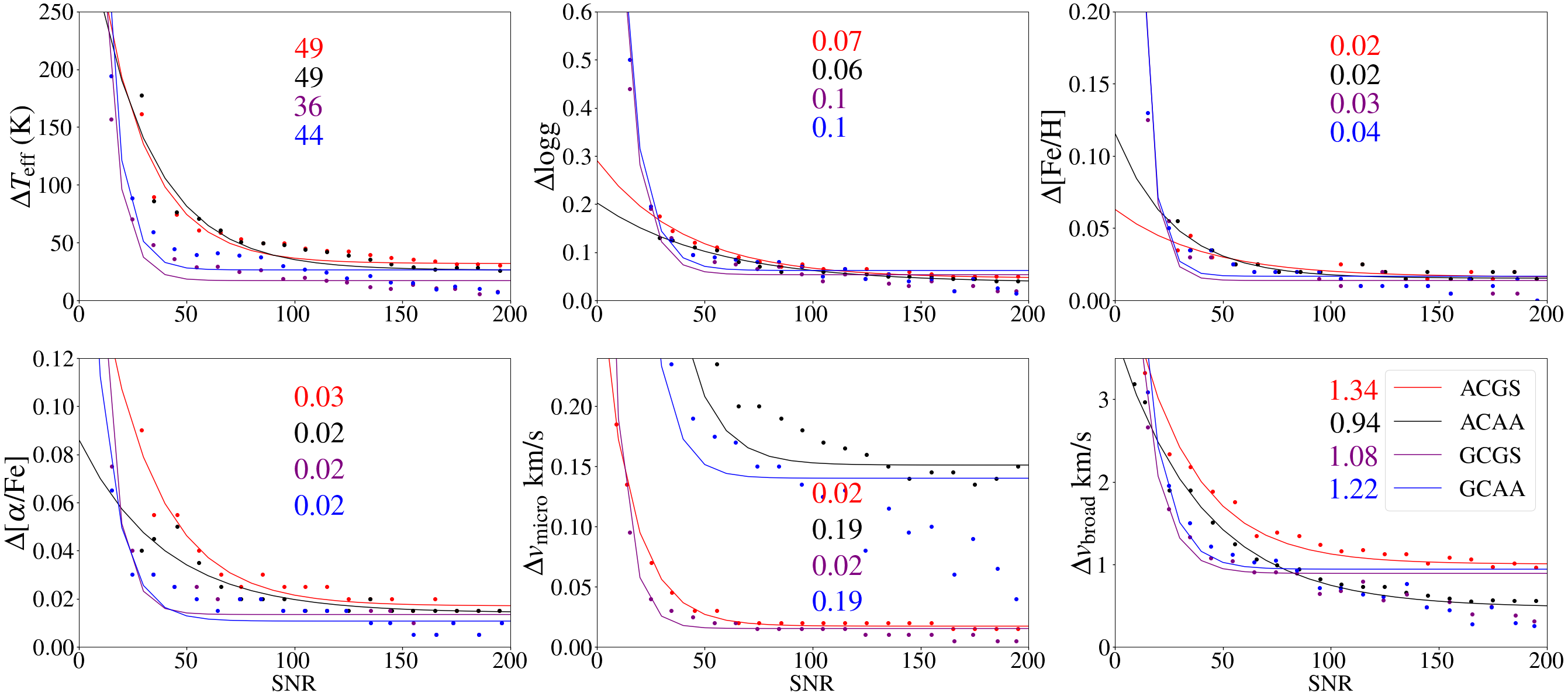}
    \caption{Exponential fits to the percentile differences (mid value of 84$^{th}$-16$^{th}$ percentile plotted as filled circles) of the \textit{Cannon} estimates of labels between high SNR and low SNR spectra among multiple visits of same stars (repeat observations) in ACGS (red), ACAA (black), GCGS (purple) and GCAA (black) binned as a function of the SNR of the lower SNR spectra. The precision for all labels except $v_{\rm mciro}$ flattens out at approximately SNR $>$ 40 when spectra belong to GALAH (GCGS and GCAA) and SNR $>$ 80 when it is APOGEE spectra (ACGS and ACAA).  }  
    \label{fig:precision_all}
\end{figure*}

\subsection{Error and precision}
\label{sec:Err}

The \textit{Cannon} provides covariance errors that are very small and likely to be underestimated \citep{Casey:2017}.
In the following subsections, we explain precision and precision estimation using repeat observations in each survey and from the label difference of stars in the training set as a function of signal to noise ratio.

\subsubsection{Error}
\label{sec:Err_Repeat}

We use the \textit{Cannon} results for repeat observations in each survey to determine the factor by which covariance errors for each label have to be scaled in order to get a reasonable error estimate. In APOGEE, there are 28,570 groups of repeated observations with the group size varying from 2 to 14. Meanwhile, there are 47,189 groups of repeated observations of sizes 2 to 15.

We calculate the pair-wise differences between the labels we derived from multiple visits over the quadrature sum of their formal covariance errors. For each label, \textit{l}, we introduce a scaling factor, $\rm f$, as shown below:

\begin{equation}
    \delta l = \frac{l_{1} - l_{2}}{\sqrt{(\sigma_{l1}^{2} + \sigma_{l2}^{2}) . f^{2}}}
\label{eq:err}
\end{equation}

Under the ideal assumption that the derived labels are unbiased and the \textit{Cannon} covariance errors are correct, we expect $\delta$\textit{l} (without scaling factor) to follow a normal distribution with zero mean and variance of unity. The resulting $\delta$\textit{l} distribution is found to be significantly different from the expected normal distribution for all labels except [$\alpha$/Fe] for \textit{Cannon} estimates in ACGS, GCGS, GCAA and ACAA. This indicates that covariance errors obtained from \textit{Cannon} are not correct. Hence, we proceed to vary $\rm f$ in Equation\ref{eq:err} until variance of $\delta$\textit{l} distribution tend to reach unity. 

 In Table~\ref{table:ErrRepeat}, we list the scaling factor for each label estimated from repeat observations for each catalogue. In all cases, scaling factor for [$\alpha$/Fe] are close to 1, which can be attributed to the fact that the \textit{Cannon} covariance error for [$\alpha$/Fe] is indicative of the actual error.

\subsubsection{Precision}
\label{sec:Precision_Repeat}

We further use repeat observations in APOGEE and GALAH to determine the label precision in the \textit{Cannon} estimates as a function of SNR. For this, we calculate the differences between the \textit{Cannon} estimates from the most high SNR spectrum and those from lower SNR spectra (from multiple visits) of the same star. These differences are then binned by SNR of the lower SNR spectra with the percentile difference (mid value of 84$^{th}$-16$^{th}$ percentile) in each bin denoting the precision achieved in the respective SNR range. 

In Figure~\ref{fig:precision_all}, we show exponential fits to the percentile differences (shown as filled circles) as a function of SNR using the repeat observations in all 4 cases : ACGS, GCGS, GCAA and ACAA, as indicated by the color of the lines in the plots. The precision for all labels except $v_{\rm mciro}$ tends to vary exponentially as a function of SNR depending on the survey to which the spectra of repeat observations belong. Precision estimated from APOGEE spectra (ACGS and ACAA) tends to flatten out at SNR $>$ 80 while precision estimated from GALAH spectra follow the same trend around SNR $>$ 40. Thus when using combined catalogues (ACGS+GCGS or GCAA+ACAA), it is advisable to choose stars with SNR $>$ 40 for GALAH spectra and SNR $>$ 80 for APOGEE spectra to make sure that the parameters are of the same precision scale. In the case of $v_{\rm mciro}$, precision in APOGEE scaled cases (GCAA and ACAA) follow similar trend independent of the spectra the label is inferred from. Also, higher precision is achieved for $v_{\rm mciro}$ in GALAH scaled cases independent of the spectra. Thus, in the case of $v_{\rm mciro}$, we infer that the \textit{Cannon} models are unable to find strong correlations for APOGEE values with either spectra. As mentioned in Section~\ref{sec:trset}, this may be the result of different methods employed to determine $v_{\rm mciro}$ in APOGEE and GALAH (also see Sections~\ref{sec:Cross-Validation} and ~\ref{app:CV}).

We also list the precision at SNR $>$ 80 (for ACGS and ACAA) and SNR $>$ 40 (for GCAA and GCGS) in Figure~\ref{fig:precision_all}. We find higher precision for T$_\mathrm{eff}$ and $\log g$ when they are estimated from GALAH spectra and APOGEE spectra respectively. Meanwhile, similar high precision ($\sim$ 0.02-0.04 dex) is obtained in all cases for [Fe/H] and [$\alpha$/Fe].




We find that the rescaled \textit{Cannon} covariance errors and precision estimates closely follow each other as a function of SNR. Hence, we take the maximum value from among the rescaled \textit{Cannon} covariance errors and precision estimate at the respective SNR as the final error for each label in all cases. This is published in the final catalogues.

\section{Validation}
\label{sec:valid}

We combine \textit{APOGEE Cannon GALAH SME} (ACGS) and \textit{GALAH Cannon GALAH SME} (GCGS) to construct the GALAH scaled catalogue. Similarly we combine \textit{GALAH Cannon APOGEE ASPCAP} (GCAA) and \textit{APOGEE Cannon APOGEE ASPCAP} (ACAA) to construct the APOGEE scaled catalogue. Both the catalogues provide stellar parameters, metallicity and global alpha abundances for 1,013,835 stars corresponding to the sum of number of unique stellar spectra in each survey (437,445 in APOGEE and 576,390 in GALAH). After applying the flags as mentioned in the Section~\ref{sec:flag}, we are left with slightly less than 50$\%$ of the total number of stars in the catalogues, but with good quality spectra and the survey parameters on a common scale in each catalogue.


In this section, we carry out external astrophysical validation of the \textit{Cannon} estimates to investigate how well the \textit{Cannon} is able to learn from GALAH and APOGEE labels, along with their respective spectra, to produce the GALAH and APOGEE scaled catalogues. To do this though, we have only a limited number of stars from other high resolution spectroscopic studies that have been observed by APOGEE and/or GALAH (e.g. \citealt{Reddy:2003,Reddy:2006,Bensby:2014} etc). So we rely on open and globular clusters, which contain stars which are born from the same parental molecular cloud and are expected to have similar ages. Hence we expect the stars in open and globular clusters to follow the same theoretical isochrone track in the Kiel diagram and to have similar metallicities and abundances.

\subsection{Astrophysical validation}
\label{sec:astrovalid}

In this section, we investigate the astrophysical validity of the stellar parameters in our catalogues. We start by searching for members of previously observed, well studied open and globular clusters in our combined GALAH scaled and APOGEE scaled catalogues. We then check the consistency of our \textit{Cannon} estimated metallicities of these member stars with that in APOGEE and/or GALAH as well as with that in the literature. 

\begin{table*}
\caption{Median metallicity and standard deviation estimated for members of 5 open clusters from the catalogue of \protect\cite{Spina:2021} cross matched with the catalogues in this work. The names of clusters are listed in the first column, metallicity of the cluster from high resolution study \citep{Heiter:2014} in second column, the abbreviations for the catalogues (see Section~\ref{sec:test}) with the number of stars from each catalogue in brackets in third column, \textit{Cannon} metallicity estimates from each catalogue in fourth column and survey pipeline metallicity estimates corresponding to the spectra mentioned (in the catalogue name) in the same row in the last column.  }\label{table:oc}
\begin{tabular}{l c c c c c }
\hline
\hline
 Cluster & [Fe/H]$_{\rm literature}$ & Catalogue (No.)  & [Fe/H]$_{\rm Cannon}$  & [Fe/H]$_{\rm Survey}$   \\
   & (median, std dev) &   & (median, std dev) & (median, std dev) &    \\
\hline
\multirow{4}{*}{IC 4665} & \multirow{4}{*}{-0.03, 0.04} & GCGS (13)  & -0.05, 0.13 & -0.03, 0.10  \\
                                                    &  &  ACGS (0) & --  &  --   \\  
                                                    &  &  GCAA (13)  & 0.03, 0.10 & -0.03, 0.10   \\
                                                    &  &  ACAA (0)   & -- &   --     \\                        
\hline
\multirow{4}{*}{Melotte 22} & \multirow{4}{*}{-0.01, 0.05} & GCGS (36)   & -0.05, 0.15 & -0.03, 0.09 \\
                                                        &  & ACGS (62)  &  0.00, 0.10 &  -0.01, 0.04     \\  
                                                        &  & GCAA (36)  & 0.00, 0.05 & -0.03, 0.09   \\
                                                        &  & ACAA (62)   & 0.00, 0.04 &   -0.01, 0.04    \\
\hline
\multirow{4}{*}{Blanco 1} & \multirow{4}{*}{0.03, 0.07} & GCGS (35)  & 0.0, 0.09 & -0.05, 0.07  \\
                                                    &   & ACGS (0) &  -- &  --   \\  %
                                                    &   & GCAA (35)  & -0.01, 0.06 & -0.05, 0.07  \\  
                                                    &  &  ACAA (0)  & -- &   --    \\  
\hline
\multirow{4}{*}{Ruprecht 147} & \multirow{4}{*}{0.16, 0.08} & GCGS (16)   & 0.05, 0.06 & 0.03, 0.06 \\
                                                        &  &  ACGS (29) & 0.07, 0.08  &  0.12, 0.03     \\  
                                                        &  &  GCAA (16)  & 0.11, 0.07 & 0.03, 0.06  \\
                                                        &  &  ACAA (29)  & 0.10, 0.05 &   0.12, 0.03     \\      
\hline
\multirow{4}{*}{NGC 2682} & \multirow{4}{*}{0.0, 0.06} & GCGS (85)   & -0.05, 0.09 & -0.06, 0.07 \\
                                                    &  & ACGS (109)  & -0.06, 0.07  &  0.00, 0.03    \\  
                                                    &  & GCAA (85)  & -0.02, 0.07 & -0.06, 0.07   \\
                                                    &  & ACAA (109)  & 0.00, 0.05 &   0.00, 0.03     \\                
             
\hline
\hline
\end{tabular}
\end{table*}

\begin{figure*}
	\includegraphics[width=\textwidth]{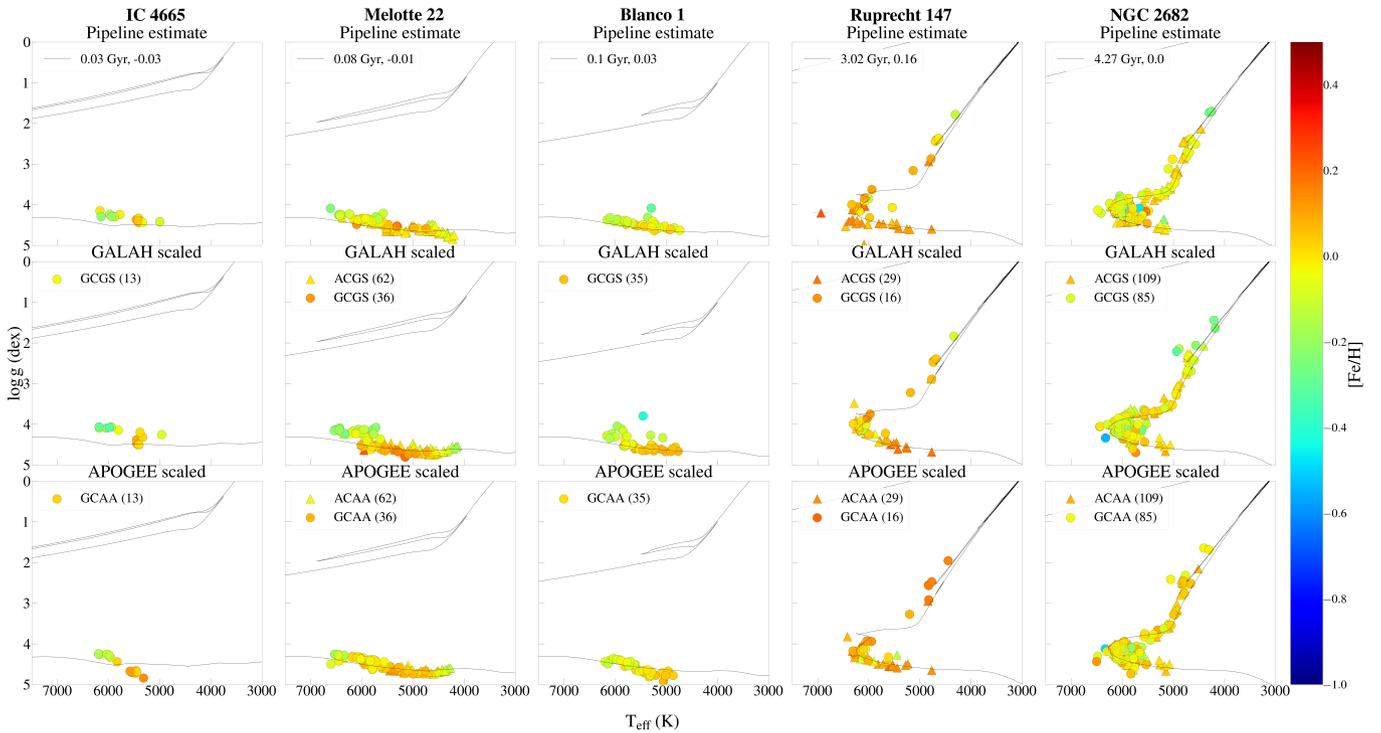}
    \caption{Kiel diagram for members of 5 open clusters from the catalogue of \protect\cite{Spina:2021} color coded by their metallicities. Clusters are arranged from left to right in increasing order of their ages from literature. We show survey pipeline (GALAH SME and APOGEE ASPCAP) estimates in the top row, GALAH scaled values (ACGS and GCGS) in the middle row and APOGEE scaled values (GCAA and ACAA) in the bottom row. PARSEC isochrones \citep{Bressan:2012} based on cluster ages and metallicities adopted from literature (age: \protect\citealt{Cantat-Gaudin:2020}, metallicity: \protect\citealt{Heiter:2014}) are over plotted (black line) in each panel. The values estimated from APOGEE spectra (ASPCAP as well as \textit{Cannon} estimates) are indicated as triangle markers, while those from GALAH spectra (SME as well as \textit{Cannon} estimates) are indicated as circle markers.}
    \label{fig:OC}
\end{figure*}

\begin{table*}
\caption{Median metallicity and standard deviation estimated for members of 5 globular clusters from the latest Gaia EDR3 globular cluster catalogue \citep{Vasiliev:2021} cross matched with the catalogues in this work. The names of clusters are listed in the first column, metallicity of the cluster from literature \citep{VandenBerg:2013} in second column, the abbreviations for the catalogues (see Section~\ref{sec:test}) with the number of stars from each catalogue in brackets in third column, \textit{Cannon} metallicity estimates from each catalogue in fourth column and survey pipeline metallicity estimates corresponding to the spectra mentioned (in the catalogue name) in the same row in the last column. }\label{table:gc}
\begin{tabular}{l c c c c c }
\hline
\hline
 Cluster & [Fe/H]$_{\rm literature}$ &  Catalogue (No.)  & [Fe/H]$_{\rm Cannon}$  & [Fe/H]$_{\rm Survey}$  \\
    &  &   & (median, std dev) & (median, std dev) &    \\
\hline
\hline
\multirow{4}{*}{NGC 104} & \multirow{4}{*}{-0.76} & GCGS (181)  & -0.66, 0.08 & -0.73, 0.07\\
                                     &   & ACGS (90) & -0.73, 0.09  &  -0.71, 0.04    \\  
                                    &    & GCAA (181)  & -0.67, 0.07 & -0.73, 0.07   \\
                                     &   & ACAA (90) & -0.76, 0.04 &  -0.71, 0.04    \\
\hline
\multirow{4}{*}{NGC 2808} & \multirow{4}{*}{-1.18} & GCGS (0)  & -- & -- \\
                                     &   & ACGS (31) & -1.10, 0.06  &  -1.10, 0.05   \\  
                                     &    & GCAA (0)  & -- & --  \\
                                     &   & ACAA (31) & -1.12, 0.05 &  -1.10, 0.05   \\
\hline
\multirow{4}{*}{NGC 6121} & \multirow{4}{*}{-1.18} & GCGS (0) & -- & -- \\
                                     &   & ACGS (72) & -1.03, 0.04  &  -1.01, 0.04   \\  
                                     &    & GCAA (0)  & -- & --   \\
                                     &   & ACAA (72) & -1.07, 0.03 &  -1.01, 0.04    \\
\hline
\multirow{4}{*}{NGC 288} & \multirow{4}{*}{-1.32} & GCGS (19)  & -1.12, 0.05 & -1.07, 0.04 \\
                                     &   & ACGS (33) & -1.24, 0.06  &  -1.24, 0.04     \\  
                                     &    & GCAA (19)  & -1.26, 0.09 & -1.07, 0.04  \\
                                     &   &  ACAA (33) & -1.26, 0.07 &  -1.24, 0.04    \\
\hline
\multirow{4}{*}{NGC 6809} & \multirow{4}{*}{-1.93} & GCGS (0)  & -- & --\\
                                     &   & ACGS (31) & -1.67, 0.04  &  -1.67, 0.07   \\  
                                    &    & GCAA (0)  & -- & --   \\
                                     &   & ACAA (31) & -1.71, 0.06 &  -1.67, 0.07    \\
\hline
\hline
\end{tabular}
\end{table*}

\begin{figure*}
	\includegraphics[width=\textwidth]{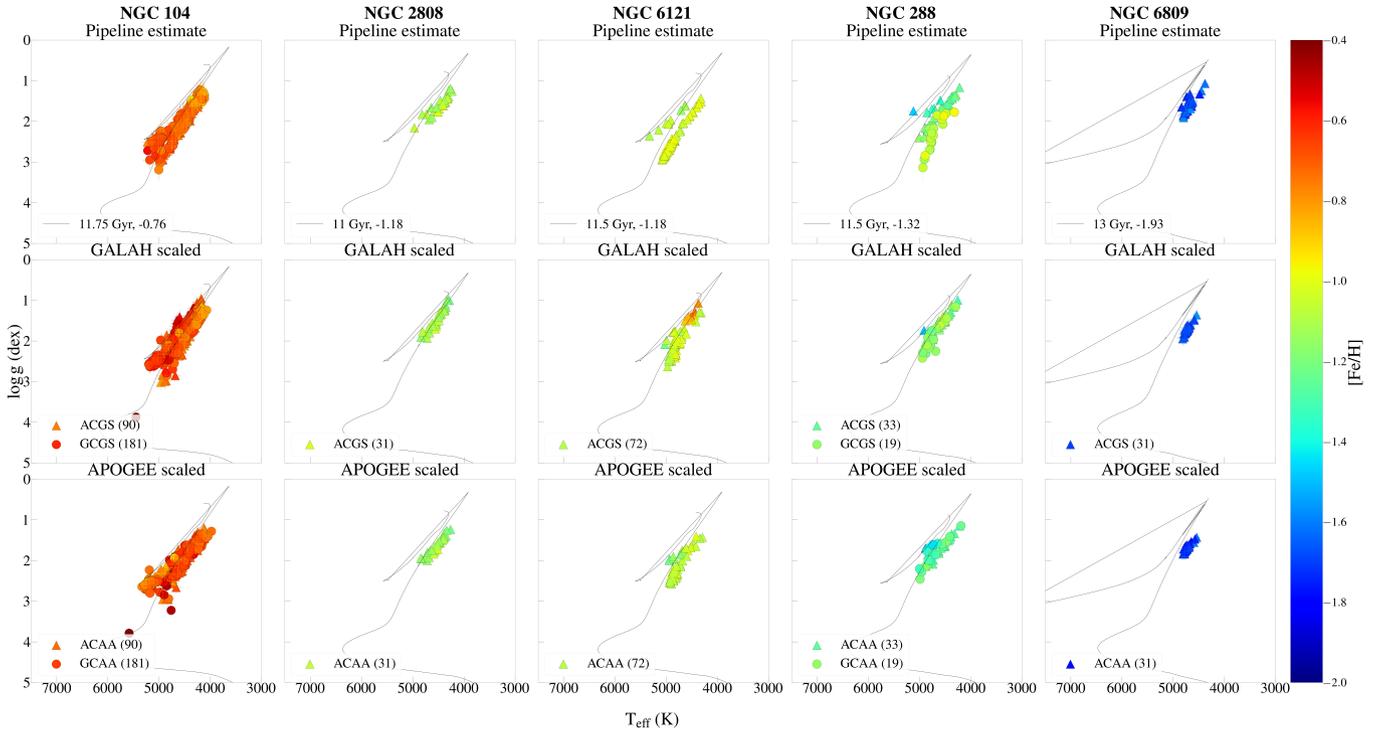}
    \caption{ Kiel diagram for members of 5 globular clusters from the latest Gaia EDR3 globular cluster catalogue \citep{Vasiliev:2021} color coded by their metallicities. Clusters are arranged from left to right in decreasing order of their median metallicities from literature. We show survey pipeline (GALAH SME and APOGEE ASPCAP) estimates in the top row, GALAH scaled values (ACGS and GCGS) in the middle row and APOGEE scaled values (GCAA and ACAA) in the bottom row. PARSEC isochrones \citep{Bressan:2012} based on cluster ages and metallicities adopted from literature \citep{VandenBerg:2013} are over plotted (black line) in each panel. The values estimated from APOGEE spectra (ASPCAP as well as \textit{Cannon} estimates) are indicated as triangle markers, while those from GALAH spectra (SME as well as \textit{Cannon} estimates) are indicated as circle markers. }
    \label{fig:GC}
\end{figure*}

\subsubsection{Open Clusters}
\label{sec:oc}

We cross match our catalogues with the open cluster member catalogue provided by \cite{Spina:2021} for 205 clusters observed by at least one of the survey, APOGEE (DR16) or GALAH (DR3). From our GALAH scaled and APOGEE scaled catalogues, we choose stars with \textit{flag$\_$Cannon$\_$dist}, \textit{flag$\_$survey}, \textit{flag$\_$spectra} set to 0, SNR cuts as in the training set and remove stars flagged by both surveys (\textit{flag$\_$sp$\_$aspcap}). From the open cluster catalogue, we choose member stars with membership probability, P$\_$mem, $>$ 0.5. We identify 113 open clusters with at least one member in our catalogues which satisfy these conditions. For our investigation, we choose 5 open clusters: Blanco 1, IC 4665, Melotte 22, NGC 2682 and Ruprecht 147.

We show Kiel diagrams for members of the above mentioned clusters color coded by their metallicities in Figure~\ref{fig:OC}. The plots for clusters are arranged from left to right in increasing order of their ages from literature \citep{Cantat-Gaudin:2020}, with the top row showing survey pipeline (SME/ASPCAP) estimates, middle and bottom row showing  GALAH scaled and APOGEE scaled values respectively for same stars. We also overplot PARSEC isochrones \citep{Bressan:2012} in each panel where the isochrone ages are from \cite{Cantat-Gaudin:2020} and metallicities are adopted from \cite{Heiter:2014}. For each cluster, we separate out the spectra of stars from each survey (indicated by different symbols in the Figure~\ref{fig:OC}) and list median metallicities and standard deviations of GALAH scaled and APOGEE scaled cases for their respective spectra in Table~\ref{table:oc}. In addition, we also list the median and standard deviation values of the pipeline metallicity estimates, enabling us to compare and quantify the similarity of \textit{Cannon} estimates to the respective survey pipeline estimates. 

For Blanco 1 and IC 4665, there are no APOGEE stars in our catalogues after quality cuts. Except for IC 4665, member stars from rest of the clusters consistently lie on the theoretical isochrone track defined by respective age and metallicity from literature, for both \textit{Cannon} estimates as well as individual survey pipeline estimates. In IC 4665, GALAH SME $\log g$ estimates also show deviation from the track at higher T$_\mathrm{eff}$ for a few stars in the GALAH scaled case, which is more pronounced in GALAH scaled case at T$_\mathrm{eff}$ $>$ 6000 K. This is not unexpected since there are fewer lines in the GALAH wavelength windows for these hotter stars. Similar deviations from tracks or scatter in $\log g$ are evident in the case of other two young clusters, Melotte 22 and Blanco 1, pointing out the problems in the GALAH SME estimates and spectra of hot dwarf stars.  

Within the standard deviations, GALAH scaled and APOGEE scaled median metallicity values are consistent with that from the median survey pipeline estimates of stars in GALAH and APOGEE surveys and with the literature values (see Table~\ref{table:oc}). At the same time, we see higher values of the standard deviations in GALAH scaled metallicities compared to APOGEE scaled cases. This is expected since GALAH SME estimates also show higher values of standard deviation compared to APOGEE ASPCAP values for all clusters. In the case of Ruprecht 147, GALAH scaled median metallicities from APOGEE spectra (ACGS) are consistent with GALAH SME median metallicity. Similarly, APOGEE scaled median metallicities estimated from GALAH spectra (GCAA) are consistent with the more metal rich APOGEE ASPCAP median metallicity estimate. These are clear indications of the ability of the \textit{Cannon} to carry out cross survey scaling.

\subsubsection{Globular Clusters}
\label{sec:gc}

Unlike in the case of open clusters, there are no combined catalogues of globular clusters in APOGEE and GALAH. So we cross match our catalogues with the latest Gaia EDR3 globular cluster catalogue \citep{Vasiliev:2021}. As in the case of open clusters, we choose stars from our catalogues with \textit{flag$\_$Cannon$\_$dist}, \textit{flag$\_$survey}, \textit{flag$\_$spectra} set to 0, SNR cuts as in the training set, globular cluster member stars with membership probability $>$ 0.5 and remove stars flagged by both surveys (\textit{flag$\_$sp$\_$aspcap}). This results in nearly 28 globular cluster populations with only two of them having spectra in both APOGEE and GALAH. For our investigation, we choose 5 globular clusters : NGC 104 (47 Tuc), NGC 2808, NGC 288, NGC 6121 and NGC 6809.

We show Kiel diagrams for members of the above mentioned clusters color coded by their metallicities in Figure~\ref{fig:GC}. The plots for clusters are arranged from left to right in decreasing order of their metallicities from literature \citep{VandenBerg:2013}, with the top row showing survey pipeline (SME/ASPCAP) estimates, middle and bottom row showing  GALAH scaled and APOGEE scaled values respectively for same stars. We also overplot PARSEC isochrones \citep{Bressan:2012} in all panels where the isochrone ages and metallicities of these clusters are from \cite{VandenBerg:2013}. In table~\ref{table:gc}, we list the median metallicities and standard deviations of GALAH scaled and APOGEE scaled cases for their respective spectra as well as from survey pipelines (SME/ASPCAP), in addition to respective cluster metallcities from \cite{VandenBerg:2013}. 

From the plots, we find \textit{Cannon} and survey pipeline estimates to follow theoretical isochrone tracks in the case of NGC 104. For NGC 2808, NGC 6121 and NGC 288, survey pipeline estimates are systematically offset (lower T$_\mathrm{eff}$) from the respective theoretical isochrone tracks while the \textit{Cannon} estimates are closer to them. This is a clear indication of the quality of \textit{Cannon} estimates compared to the survey pipeline estimates and shows the ability of the \textit{Cannon} models to correct the shift in pipeline estimates by learning it from field stars that are dominant in the training set. For the most metal poor cluster in our sample, NGC 6809, both the \textit{Cannon} and survey pipeline estimates are offset from the stellar track. For this cluster, the median metallicity from the literature is $\sim$0.25 dex more metal poor than those from the \textit{Cannon} as well as survey pipeline estimates. This is most likely the reason for the shift in \textit{Cannon} and pipeline estimates compared to the theoretical isochrone track. On the other hand, during cross-validation of the training set, we have seen that the \textit{Cannon} estimates for stars with metallicities lower than -1.5 dex may be wrong due to them being scarcely represented in the training set and values close to the training set boundary. But consistency of \textit{Cannon} estimates with the survey pipeline estimates gives strength to the validity of the \textit{Cannon} estimates in this case.

For all clusters except NGC 288 and NGC 104, median metallicities from the \textit{Cannon} are consistent with the median metallicity values from survey pipelines. In NGC 288, median metallicity from GALAH SME (-1.07 dex) is more metal rich than GALAH scaled (-1.12 dex) and APOGEE scaled (-1.26 dex) values from GALAH spectra for same set of stars, while we find consistent median metallicities from APOGEE ASPCAP and \textit{Cannon} estimates from APOGEE spectra. This means that for NGC 288 members, metallicity information estimated from GALAH spectra with the \textit{Cannon} model trained on GALAH spectra using APOGEE ASPCAP labels (GCAA) results in more accurate (and closer to literature estimate) metallicity values compared to GALAH SME values. Similarly, the \textit{Cannon} model trained on APOGEE spectra using GALAH SME labels (ACGS) is able to derive consistent metallicity information from APOGEE spectra, while slight decrease in median metallicity compared to GALAH SME estimate is seen in the case of GCGS. This points out that the GALAH SME metallicity estimates could be wrong, which is confirmed in the next section where we find significant systematic differences in stellar parameters for these stars between APOGEE and GALAH.

In NGC 104, median metallicity estimated from \textit{Cannon} models trained on GALAH spectra using GALAH SME (GCGS) and APOGEE ASPCAP (GCAA) labels are more metal rich compared to the pipeline estimates as well as literature value. On further investigation by comparing GALAH spectra of NGC 104 members with similar T$_\mathrm{eff}$, $\log g$ and [Fe/H] from GALAH SME, we found differences in spectra of stars with more metal rich \textit{Cannon} estimates. This difference is found to be due to enhanced N abundance in stars with more metal rich \textit{Cannon} estimates resulting in strong CN bands as shown in Figures~\ref{fig:NGC104_cont} and \ref{fig:NGC104_strength}. Thus these pervasive CN lines may blend with atomic features, resulting in higher [Fe/H] estimates by the \textit{Cannon} models for these stars. These stars are also found to belong to the second generation population in NGC 104 based on their enhanced N, Na, Al and low O, Mg abundances from GALAH SME. At the same time, majority of NGC 104 members in our APOGEE sample (ACGS and ACAA) belongs to the first generation population based on their ASPCAP light element abundances. Hence we do not see similar difference in [Fe/H] \textit{Cannon} estimates compared to the pipeline estimates using APOGEE spectra (ACGS and ACAA).

For all clusters excluding young open clusters, we find a mean scatter of 0.06 dex in APOGEE scaled metallicity estimates and slightly higher scatter of 0.065 dex in GALAH scaled metallicity estimates. we find a smaller mean scatter of 0.03 dex in both APOGEE and GALAH scaled alpha abundance estimates. Overall, we find the \textit{Cannon} metallicity estimates to be consistent with that in literature and follow expected theoretical isochrone tracks in the Kiel diagram, showing the quality of our GALAH and APOGEE scaled catalogues. 

\begin{figure*}
	\includegraphics[width=\textwidth]{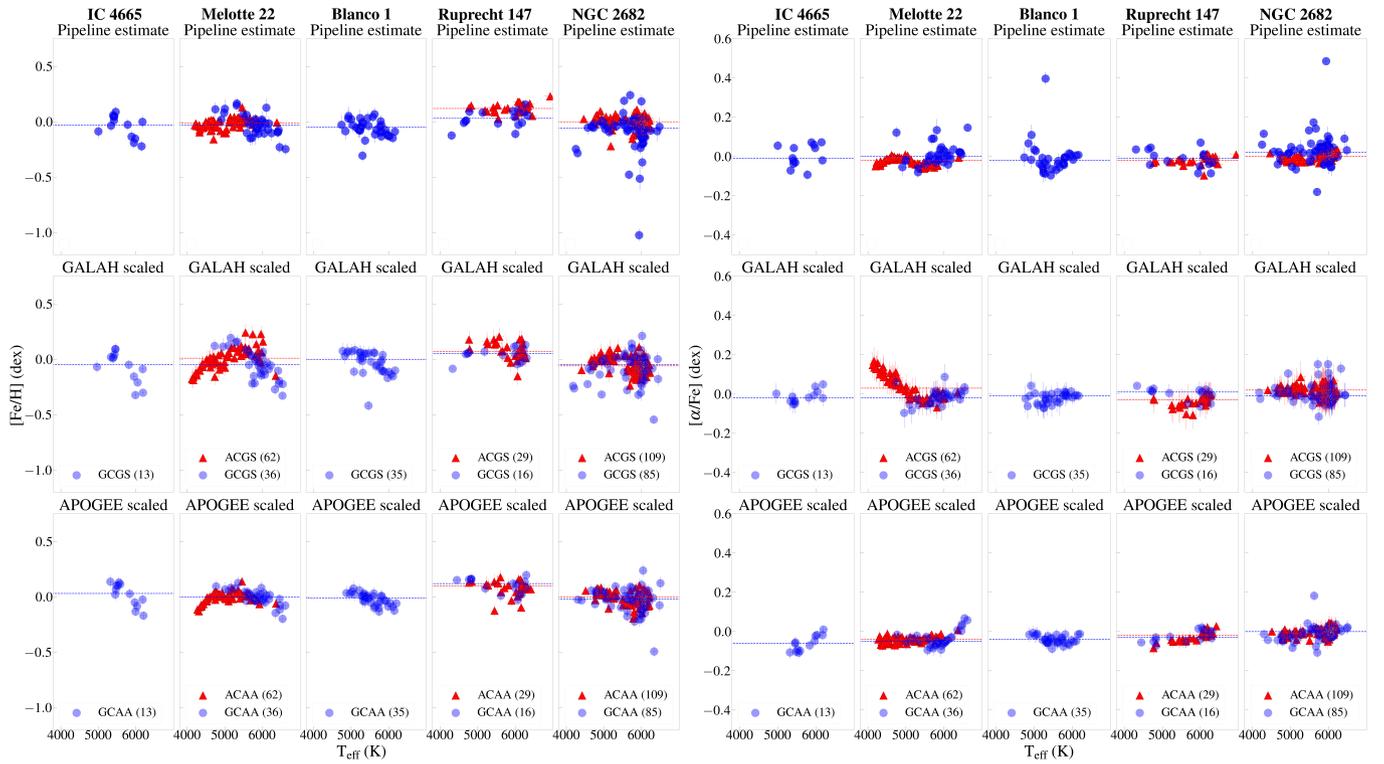}
    \caption{[Fe/H] vs T$_\mathrm{eff}$(left 5 columns) and [$\alpha$/Fe] vs T$_\mathrm{eff}$ (right 5 columns) plots for 5 open clusters. We show survey pipeline (GALAH SME and APOGEE ASPCAP) estimates in the top row, GALAH scaled values (ACGS and GCGS) in the middle row and APOGEE scaled values (GCAA and ACAA) in the bottom row. Estimates from GALAH spectra are indicated in blue circles and from APOGEE spectra in red triangles, with the median metallicity and [$\alpha$/Fe] values in each case indicated with respectively colored dashed lines. }
    \label{fig:OC_teff}
\end{figure*}

\begin{figure*}
	\includegraphics[width=\textwidth]{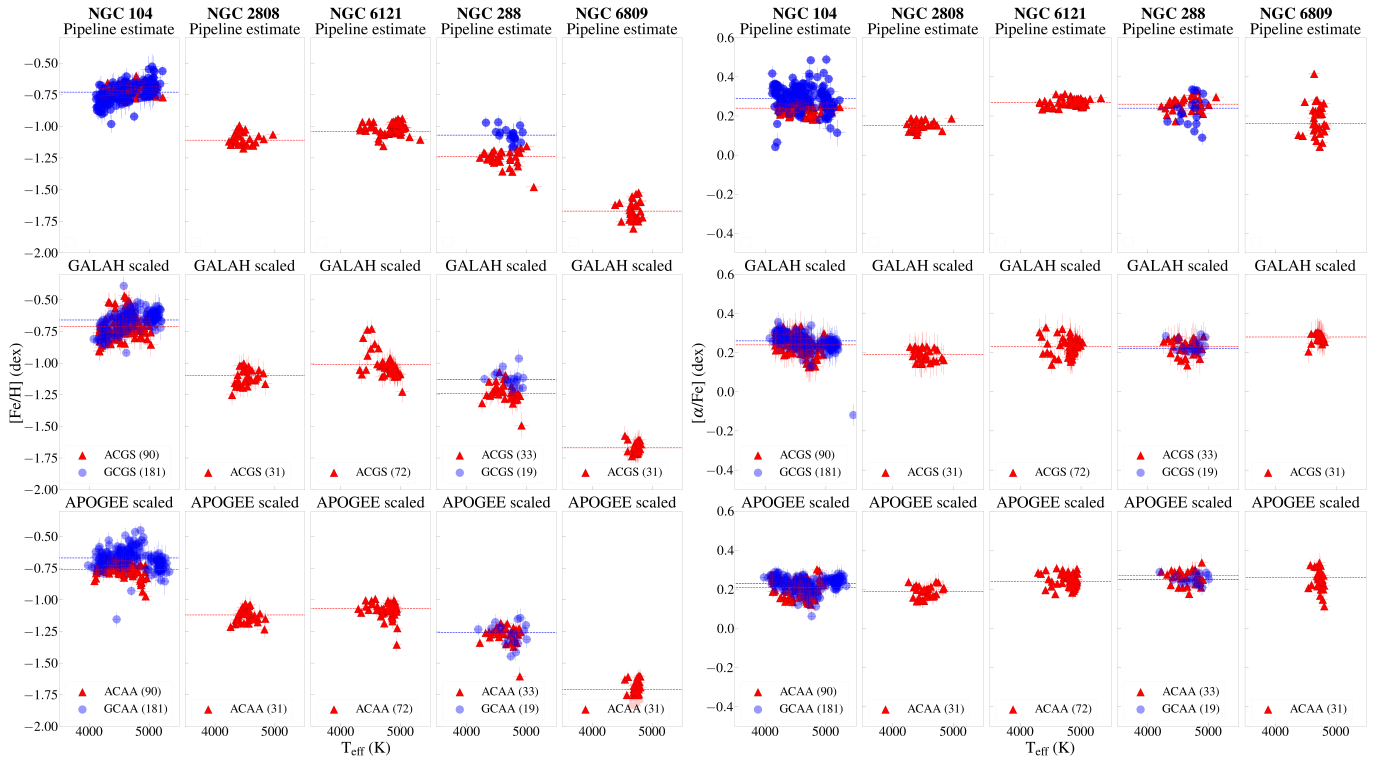}
    \caption{[Fe/H] vs T$_\mathrm{eff}$ (left 5 columns) and [$\alpha$/Fe] vs T$_\mathrm{eff}$ (right 5 columns) plots for 5 globular clusters in the same format as in Figure~\ref{fig:OC_teff}}
    \label{fig:GC_teff}
\end{figure*}

\begin{figure}
\hspace*{-0.5cm} 
	\includegraphics[width=0.5\textwidth]{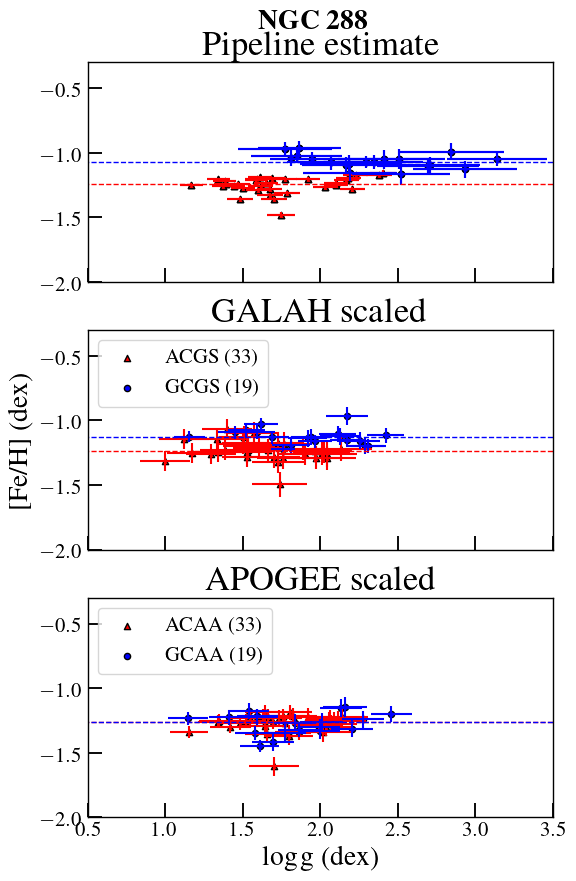}
    \caption{[Fe/H] vs $\log g$ for NGC 288 with survey pipeline (GALAH SME and APOGEE ASPCAP) estimates in the top row, GALAH scaled values (ACGS and GCGS) in the middle row and APOGEE scaled values (GCAA and ACAA) in the bottom row. Estimates from GALAH spectra are indicated in blue circles and from APOGEE spectra in red triangles, with the median metallicity in each case indicated with respectively colored dashed lines.}
    \label{fig:GC_NGC288}
\end{figure}

\subsubsection{Chemical trends in clusters}
\label{sec:trends}

Though the median metallicities from the \textit{Cannon} estimates show good consistency with that from respective survey pipelines as well as those from the literature, it is important to check for any significant trends in chemistry (metallicity and alpha abundance) with temperature. For cluster members (both open and globular clusters), we expect such trends to be non existent owing to the chemical homogeneity of their birth cloud. At the same time, atomic diffusion can lead to differences in surface abundance values for different evolutionary phases (i.e. main sequence turn off stars and giants) of cluster members as well \citep{Dotter:2017}. In addition to atomic diffusion, \cite{Spina:2020} have shown that stellar parameters and abundances vary as a function of the stellar activity in young stars. Hence, we also have to take this into account while investigating chemical trends in clusters.

In this work, where we have used the \textit{Cannon} to estimate GALAH scaled labels from APOGEE spectra (and GALAH spectra) and APOGEE scaled labels from GALAH (and APOGEE spectra), we expect the \textit{Cannon} estimates to follow the trends we see in the case of survey labels which have been used to train the model with. Thus, we expect GALAH scaled cases (ACGS and GCGS) to exhibit similar trends as seen with GALAH SME values and APOGEE scaled cases (GCAA and ACAA) that of APOGEE ASPCAP. All figures discussed below are in the same format as in previous section (Figures~\ref{fig:OC} and \ref{fig:GC}), with the top row showing the survey pipeline (SME/ASPCAP) estimates, middle and bottom rows showing the GALAH scaled and APOGEE scaled values respectively for the same stars. Estimates from GALAH spectra are indicated in blue circles and from APOGEE spectra in red triangles, with the median metallicity values in each case indicated with respectively colored dashed lines.

In Figure~\ref{fig:OC_teff}, first 5 columns show metallicity trends of 5 open clusters with respect to T$_\mathrm{eff}$ and the next 5 columns show [$\alpha$/Fe] trends of the same clusters with respect to T$_\mathrm{eff}$. We have arranged them from left to right in increasing order of their ages, from which it is clear that chemical trends with T$_\mathrm{eff}$ are prominent for youngest clusters (IC 4665, Blanco 1 and Melotte 22). From GALAH SME estimates (first 3 columns in top row, blue circles), we see a slight negative trend for metallicity and positive trend for [$\alpha$/Fe] with increase in T$_\mathrm{eff}$. As mentioned above, this could be the effect of stellar activity in young cluster members which is expected to strengthen saturated lines, resulting in a complicated interplay between effects on T$_\mathrm{eff}$, $\log g$ and chemical abundances \citep{Spina:2020}. Among young clusters, only Melotte 22 has stars from both APOGEE and GALAH. For this cluster, GALAH scaled metallicities and alpha abundances enhance the trends imprinted from the SME pipeline compared to APOGEE scaled estimates. APOGEE ASPCAP [$\alpha$/Fe] values in Melotte 22 shows a wave like trend with T$_\mathrm{eff}$, which completely disappears in the APOGEE scaled values from APOGEE spectra. Meanwhile GALAH scaled [$\alpha$/Fe] values are higher for the same stars, a trend that is seen in GALAH SME estimates for cool stars in Blanco 1. Meanwhile, this trend is removed in the case of APOGEE and GALAH scaled [$\alpha$/Fe] values for Blanco 1. For older open clusters, Ruprecht 147 and NGC 2682, there are no significant trends seen in survey pipeline estimates and correspondingly in the \textit{Cannon} estimates, with respect to T$_\mathrm{eff}$. We also note that for NGC 2682, the median metallicity for the GALAH scaled case and APOGEE scaled case align with the respective survey pipeline estimates.

In Figure~\ref{fig:GC_teff}, first 5 columns show metallicity trends of 5 globular clusters with respect to T$_\mathrm{eff}$ and the next 5 columns show [$\alpha$/Fe] trends of the same clusters with respect to T$_\mathrm{eff}$. We have arranged them from left to right in decreasing order of their metallicities. Unlike in open clusters, there are no obvious trends with T$_\mathrm{eff}$ for survey pipeline estimates and \textit{Cannon} estimates. In all clusters, GALAH scaled estimates have larger standard deviations or scatter as listed in the Table~\ref{table:gc}. As mentioned in the previous section, for NGC 104 differences in spectra due to strong CN bands in second generation stars results in higher metallicity values estimated from GALAH spectra in both GALAH scaled and APOGEE scaled cases (see Figures~\ref{fig:NGC104_cont} and \ref{fig:NGC104_strength}). In addition to this, there are few GALAH stars for which APOGEE scaled T$_\mathrm{eff}$ is higher (by $\sim$100-150 K) than GALAH SME estimates. On further investigation, these are found to be red clump stars. In the training set, we find similar difference in T$_\mathrm{eff}$ between APOGEE ASCPAP and GALAH SME estimates for red clump stars in the same GALAH T$_\mathrm{eff}$ and $\log g$ ranges. Thus this offset is likely real that is propagated from the systematic differences in training set.

 In the case of NGC 288, GALAH SME and APOGEE ASPCAP metallicity values are different ($\sim$0.2 dex as shown in Table~\ref{table:gc}). GALAH scaled metallicity estimates for GALAH and APOGEE stars also show difference which is smaller than that between survey pipeline estimates. Meanwhile, APOGEE scaled metallicity estimates from both set of stars are consistent with the APOGEE ASPCAP estimate. At the same time, GALAH scaled and APOGEE scaled [$\alpha$/Fe] estimates are consistent for both APOGEE and GALAH stars and with respective survey pipeline estimates. In Figure~\ref{fig:GC_NGC288}, we show metallicity trends with $\log g$ for survey pipeline estimates (top panel), GALAH scaled estimates (middle panel) and APOGEE scaled estimates (bottom panel). There is a clear offset between the $\log g$ values of GALAH and APOGEE pipeline estimates (see Figure~\ref{fig:GC_NGC288}), which resulted in the shift from theoretical isochrone tracks seen in Figure~\ref{fig:GC}. This offset is found to be the result of bad parallax measurements for NGC 288 members (located at a distance of $\sim$9 kpc) leading to wrong GALAH SME $\log g$ estimates (Equation 1 in \citealt{GALAHDR3}). Interestingly, GALAH scaled and APOGEE scaled $\log g$ for GALAH stars are shifted to correct $\log g$ values (blue circles in middle and bottom panels in Figure~\ref{fig:GC_NGC288}). This could also explain the improvement in APOGEE scaled and GALAH scaled metallicity estimates for GALAH stars. Thus our \textit{Cannon} models are able to improve incorrect pipeline estimates.

Overall, we find that in most cases, GALAH scaled and APOGEE scaled metallicities and [$\alpha$/Fe] values are following the trends exhibited by the respective survey pipeline estimates. We also find GALAH scaled estimates to have comparatively higher scatter compared to APOGEE scaled values, which is expected from their respective survey pipeline values. The plots discussed above thus clearly display both qualities as well as limitations of our catalogues and show the ability of the \textit{Cannon} to carry out cross survey scaling of large data sets from high resolution spectroscopic surveys.

\section{Discussion and conclusions}
\label{sec:limitations}

\begin{figure*}
	\includegraphics[width=\textwidth]{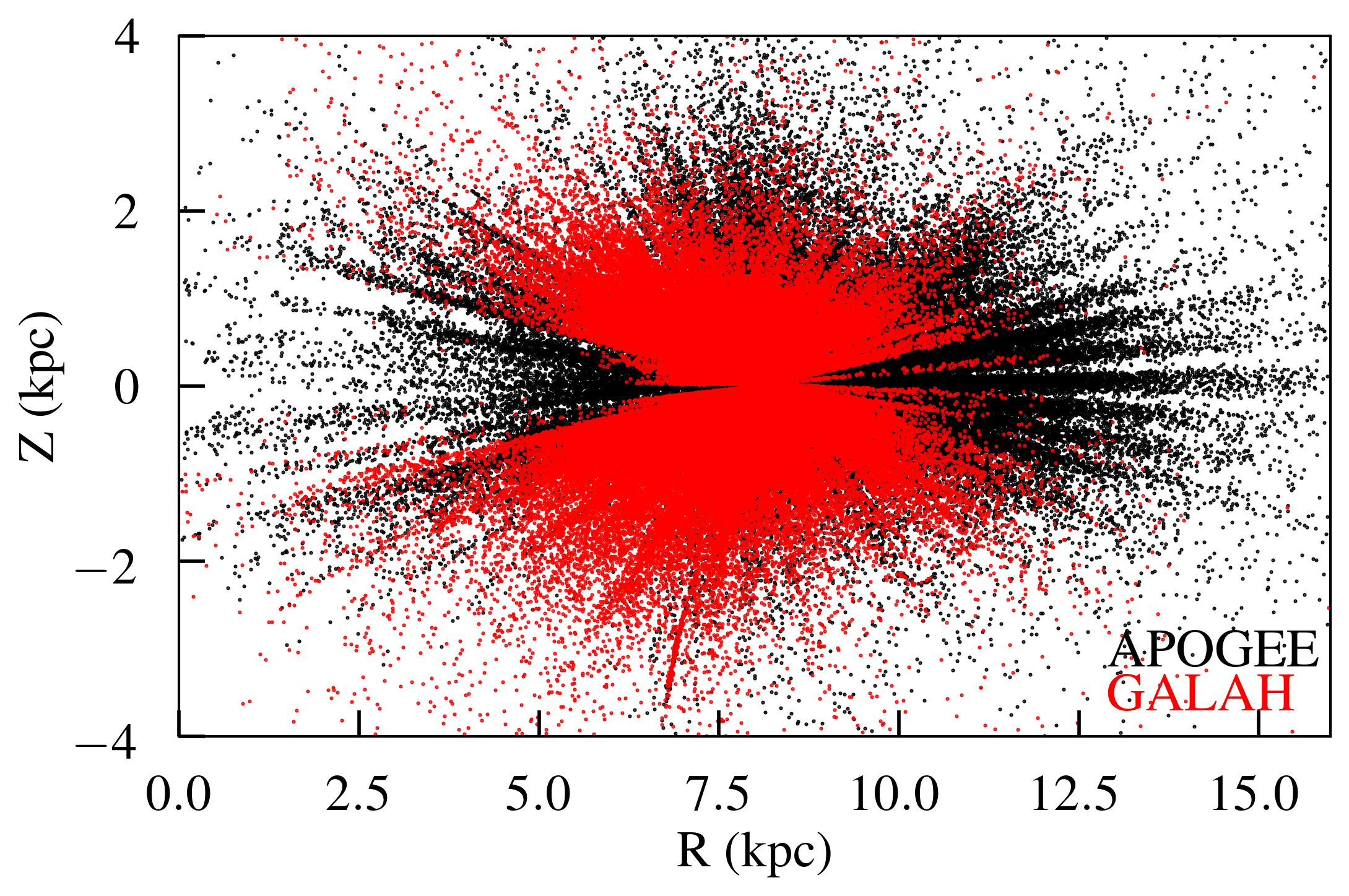}
    \caption{ Z vs R distribution of stars with valid \textit{Cannon} estimates based on flags set in both GALAH scaled (black) and APOGEE scaled (red) stellar parameter catalogues. R and Z are estimated based on the \textit{Gaia} EDR3 distances from \protect\cite{Bailer-Jones:2021}. }
    \label{fig:AGGA_RZ}
\end{figure*}

We used the data driven approach, the \textit{Cannon}, to produce two catalogues of stellar parameters and general alpha abundances, one scaled in terms of the APOGEE survey and the other in terms of the GALAH survey. We chose the training set from among $\sim$ 20,000 stars commonly observed in both the surveys, which after quality cuts (removing stars with low SNR, bad spectra, bad pipeline estimates etc) resulted in a final training set sample of 4,418 stars. We have shown the importance of cross survey scaling with plots showing the systematic differences in stellar parameters, metallicities and alpha abundances using common observed stars in the training set. While is reasonably good agreement between the two surveys (especially for T$_\mathrm{eff}$ and $\log g$), there are systematic differences in [Fe/H], [$\alpha$/Fe], $v_{\rm micro}$ and $v_{\rm broad}$ for the same stars in both surveys (Figure~\ref{fig:Sysdiff_trset_ApgGalah}). Most notably, [Fe/H] for giants and hot dwarfs are underestimated by GALAH when compared to APOGEE, while GALAH [$\alpha$/Fe] measurements are overestimated for giants. Most of these offsets could be attributed to differences in the parameter estimation methods implemented in the APOGEE and GALAH surveys. We used our final training set to train four \textit{Cannon} models in four combinations of spectra and labels, i.e., two models with spectra and labels from the different surveys respectively, and two models with spectra and labels from the same survey.

With the 12-fold cross validation of stars in the training set (see Sections~\ref{sec:Cross-Validation} and \ref{app:CV}), we have shown that the \textit{Cannon} models predict labels/parameters that are consistent with the input labels. We further demonstrate in Figures ~\ref{fig:spectra_GS} and ~\ref{fig:spectra_AA} that the model generated spectra with the \textit{Cannon} estimates are very similar (16$^{th}$ and 84$^{th}$ percentile values of residuals at majority of pixels within -0.01 and 0.01) to the respective observed survey spectra. We have also shown that pixels with higher residuals are either bad/noisy pixels or lines of elements that have not been modeled by our \textit{Cannon} models. We also demonstrate the effectiveness of our method in Figures~\ref{fig:sysdiff_ACGS_GCGS} and ~\ref{fig:sysdiff_ACAA_GCAA} where we remade the Figure~\ref{fig:Sysdiff_trset_ApgGalah} for common stars in APOGEE and GALAH but with \text{Cannon} estimates from GALAH scaled and APOGEE scaled catalogues. We show that the mean systematic difference for metallicity and alpha abundance is very close to zero with lesser scatter and no significant trends in comparison with survey pipeline estimates in Figure~\ref{fig:Sysdiff_trset_ApgGalah}.

These models are then used to estimate labels from 437,445 APOGEE and 576,390 GALAH spectra, thus providing 2 combined catalogues with stellar parameters scaled to both the APOGEE and GALAH surveys. We carried out validation of these catalogues with selected open and globular cluster members by comparing survey pipeline estimates and \textit{Cannon} estimates in kiel diagrams and chemical trend plots as shown in Figures~\ref{fig:OC}-\ref{fig:GC_teff}. In Kiel diagrams, both GALAH scaled and APOGEE scaled \textit{Cannon} estimates follow respective theoretical isochrone tracks (based on literature ages and metallicities) even when survey pipeline estimates do not (NGC 2808, NGC 6121 and NGC 288). In addition, median metallicity estimates within standard deviations are consistent with literature values thus validating our \textit{Cannon} estimates. In the case of T$_\mathrm{eff}$ trends for metallicities and alpha abundance in clusters, as expected, \textit{Cannon} estimates follow spurious trends exhibited by respective pipeline estimates to which they are scaled. Meanwhile, there are a few cases where GALAH scaled metallicities and alpha abundances enhance the trends imprinted from the SME pipeline, particularly for the very young clusters.

 Significant offsets of $\sim$0.5-1 dex between GALAH SME and APOGEE ASPCAP $\log g$ estimates for NGC 288 members (located at a distance of $\sim$9 kpc) in Figure~\ref{fig:GC_NGC288} showed that incorrect parallaxes can have an effect on $\log g$ determination in GALAH. Interestingly, $\log g$ values of GALAH stars in both GALAH scaled and APOGEE scaled cases have been shifted to the correct range. This has also led to the correction of the overestimated GALAH SME metallicities in the APOGEE scaled case and significant improvement in the GALAH scaled case. This shows the ability of the \textit{Cannon} models to correct survey pipeline estimates and illustrates the quality of our \textit{Cannon} estimates.
 
 
We have thus demonstrated the effectiveness of our method to carry out cross survey scaling and showed the quality of the resulting \textit{Cannon} estimates in both catalogues. However there are some limitations and caveats in the method that we briefly discuss below.
 
 \subsection*{Limitations in the training set}

The training set is the starting point and one of the major factors that can affect parameter estimation using data driven approaches like the \textit{Cannon}. While the training set should be representative of different populations of stars that the survey (from which the input labels are taken) observed, the labels and spectra should also be reliable and of high quality. Since we are restricted to choosing the common stars observed in both the surveys for the training set in this work, we have limited options regarding the former criterion. We have carried out quality cuts as recommended by each survey for both the labels as well as spectra, ensuring a good quality training set. Ideally, in cases where the spectra and label are from the same survey (ACAA and GCGS) one can use a larger training set. Since our goal is to put the surveys on the same scale, we choose a single training set that passes all quality criteria and have sufficient number statistics. Thus our training set is limited by the fact that not all the best quality spectra or labels in the respective catalogue are included in it, rather we select the best labels from among the common stars observed in both the surveys. Still our final training set sample has reasonably broad coverage of the parameter space of both surveys (see Figure~\ref{fig:HR_alphafe}), but one should exercise caution at the edges of the training set, in particular for metal-poor ([Fe/H]$<-0.8$) stars and for M-dwarfs, for which the \textit{Cannon} labels deviate from the one-to-one relation. We have assigned dedicated flags that will help the user to identify such stars (see Section~\ref{sec:flag}). 

\subsection*{Propagation of parameter trends}

By reproducing APOGEE scaled labels from GALAH spectra and GALAH scaled labels from APOGEE spectra, we are propagating the trends/issues in these parameters that arise from the input survey pipeline/analysis method. Such issues are evident in the labels when we compare them in Figure~\ref{fig:Sysdiff_trset_ApgGalah}. These include the lower metallicities determined for metal rich giants in GALAH and large differences in microturbulence and broadening values for stars in GALAH as well as APOGEE (see Section~\ref{sec:trset}). We show the inability of the \textit{Cannon} models to reproduce APOGEE microturbulence values in Figures~\ref{fig:CV_GalahApg} and \ref{fig:CV_GalahApg_APOGEE}. We also see the effect of this on \textit{Cannon} temperature and surface gravity estimates in the case of ACAA (see Figure~\ref{fig:CV_GalahApg_APOGEE}). When we adopted the empirical relations used in GALAH to redetermine microturbulence values for APOGEE, these trends are corrected (see Figure~\ref{fig:CV_GalahApg_APOGEE_emp}). Thus it is important to have accurate and reliable input labels in order to get better results from data driven methods and machine learning tools and future surveys should aim to achieve this.


\subsection*{Limitations in the error determination}

With repeat observations we have been able to show that the errors for the labels from the \textit{Cannon} are either incorrect or underestimated. We then estimate scaling factor for each label and rescale their \textit{Cannon} covariance errors. Using repeat observations, we also estimated precision for all labels as a function of SNR in the range of 36-50 K (T$_\mathrm{eff}$), 0.06 - 0.1 dex ($\log g$), 0.02-0.04 dex ([Fe/H]) and 0.02-0.03 dex ([$\alpha$/Fe]) for SNR $>$ 40 in GALAH and SNR $>$ 80 in APOGEE. We choose the maximum value among rescaled covariance uncertainty and precision estimate at respective SNR to be the final error for each label. Thus our errors are determined without taking into account any possible additional dependence on temperature and surface gravity as it is done in APOGEE \citep{Jonsson:2020}. Thus there is a possibility that our errors are still underestimated. \\
\\

Finally we have $\sim$ 1 million stars with APOGEE scaled and GALAH scaled parameters in both catalogues, with spatial coverage that spans inner and outer parts of the Milky Way including the thin disc, the thick disc and the halo components.The mean metallicity and alpha abundance errors is 0.07 dex and 0.06 dex in the GALAH scaled case, and 0.06 dex and 0.03 dex in the APOGEE scaled cases. Once we implement flags in both catalogues to remove stars outside the training set boundary (\textit{flag$\_$Cannon$\_$dist}), stars with bad spectra (\textit{flag$\_$spectra}), bad SME and ASPCAP labels (\textit{flag$\_$sp$\_$aspcap} and \textit{flag$\_$survey}) and good snr (\textit{flag$\_$snr}), we have $\sim$280,000 GALAH stars and $\sim$170,000 stars from APOGEE, with a spatial coverage that spans the mid plane as well as halo regions of the Milky Way as shown in Figure~\ref{fig:AGGA_RZ}. Thus we end up with less than 50$\%$ of the total number of stars in the catalogues, but with good quality spectra and the parameters on a common scale in each catalogue. This is still an impressive number of stars with good spatial coverage to study the metallicity and alpha abundance trends of the Milky Way components.

 Combining these catalogues with the latest Gaia EDR3 catalogue enables a comprehensive chemo-dynamic study of these different components that constitutes the Milky Way from the perspective of APOGEE and GALAH, separately. We have cross matched with \textit{Gaia} EDR3 and included source ids from both DR2 and DR3 along with distances estimated by \cite{Bailer-Jones:2021} in the final catalogues as listed in the table schema, Table~\ref{table:scheme}.


\subsection*{Final conclusions}

 With this work, we have demonstrated the ability of the data driven method, \textit{Cannon}, to reliably estimate respective survey scaled stellar parameters, metallicity and alpha abundance. While having a few caveats, we find that the \textit{Cannon} estimates are as good as or showed improvement over the respective survey pipeline estimates. These are encouraging signs to use currently available data driven and machine learning tools to do cross survey scaling with many more ongoing and upcoming spectroscopic surveys. 
 
 Among the limitations and caveats discussed above, the limited size and coverage of parameter space by the common stars observed in both surveys for the training set is foremost. This will hopefully improve in coming years, as both GALAH and APOGEE South continue to observe larger samples of stars going forward. Improvements in stellar model atmospheres and incorporation of 3D NLTE models in spectroscopic analysis will enable ongoing as well as upcoming ground based surveys like SDSS-V \citep{Kollmeier:2017}, 4MOST \citep{deJong:2019}, WEAVE \citep{Dalton:2018} etc., to achieve better accuracy and precision in stellar parameter and elemental abundances. Thus we will have all the necessary ingredients to construct training sets that can be used to carry out cross survey scaling which will provide catalogues of stars with consistent stellar parameters and elemental abundances with significant coverage of all components of the Galaxy.

\section*{Acknowledgements}

We thank the anonymous referee for the comments provided which considerably improved this manuscript. GN thanks Nils Ryde and Mathias Schultheis for their support.
Based on data acquired through the Australian Astronomical Observatory, under programmes: A/2013B/13 (The GALAH pilot survey); A/2014A/25, A/2015A/19, A2017A/18 (The GALAH survey phase 1), A2018 A/18 (Open clusters with HERMES), A2019A/1 (Hierarchical star formation in Ori OB1), A2019A/15 (The GALAH survey phase 2), A/2015B/19, A/2016A/22, A/2016B/10, A/2017B/16, A/2018B/15 (The HERMES-TESS program), and A/2015A/3, A/2015B/1, A/2015B/19, A/2016A/22, A/2016B/12, A/2017A/14, (The HERMES K2-follow-up program). We acknowledge the traditional owners of the land on which the AAT stands, the Gamilaraay people, and pay our respects to elders past and present.

LC is the recipient of the ARC Future Fellowship FT160100402.

Funding for the Sloan Digital Sky Survey IV has been provided by the Alfred P. Sloan Foundation, the U.S. Department of Energy Office of Science, and the Participating
Institutions. SDSS acknowledges support and resources from the Center for High-Performance Computing at the University of Utah. The SDSS web site is www.sdss.org.

This work has made use of data from the European Space Agency (ESA) mission Gaia (http://www.cosmos.esa.int/gaia), processed by the Gaia Data Processing and Analysis Consortium (DPAC, http://www.cosmos.esa.int/web/gaia/dpac/consortium). Funding for the DPAC has been provided by national institutions, in particular the institutions participating in the Gaia Multilateral Agreement.

This publication made use of NASA's Astrophysics Data System.

\section*{Data Availability}

The data underlying this article are available in the Data Central at
https://cloud.datacentral.org.au/teamdata/GALAH/public/GALAH$\_$DR3/




\bibliographystyle{mnras}
\bibliography{GalahApogee} 






\begin{appendix} 

\section{\textit{GALAH Cannon GALAH SME} (GCGS) $\&$ \textit{APOGEE Cannon APOGEE ASPCAP (ACAA)}}
\label{app:Galah} 

We use the training set to generate \textit{Cannon} model trained on GALAH (APOGEE) spectra using GALAH SME (APOGEE ASPCAP) labels, which is then used to predict stellar parameters for rest of GALAH (APOGEE) spectra. We carry out 12-fold cross validation, error estimation from repeat observations and internal validation for both cases as shown below.

\subsection{Cross-Validation}
\label{app:CV}
\begin{figure*}
	\includegraphics[width=\textwidth]{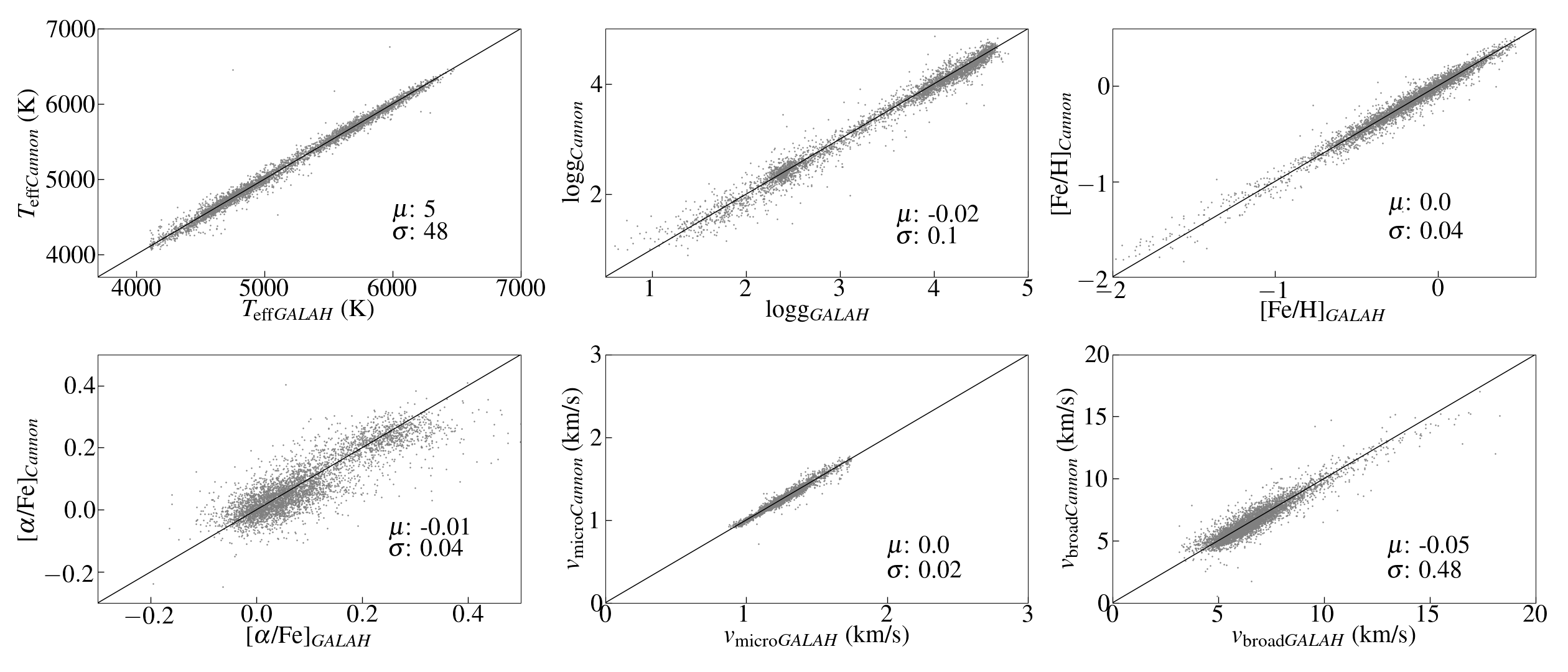}
    \caption{One-to-one relation of GALAH SME stellar labels (x-axis) in the training set versus corresponding values estimated by the \textit{Cannon} from GALAH spectra (y-axis) after carrying out a 12-fold cross-validation test. The mean and scatter (calculated as the mid value of 84$^{th}$-16$^{th}$ percentile) in each label difference is indicated on the bottom right hand side of each plot. }
    \label{fig:CV_ApgGalah_GALAH}
\end{figure*}

\begin{figure*}
	\includegraphics[width=\textwidth]{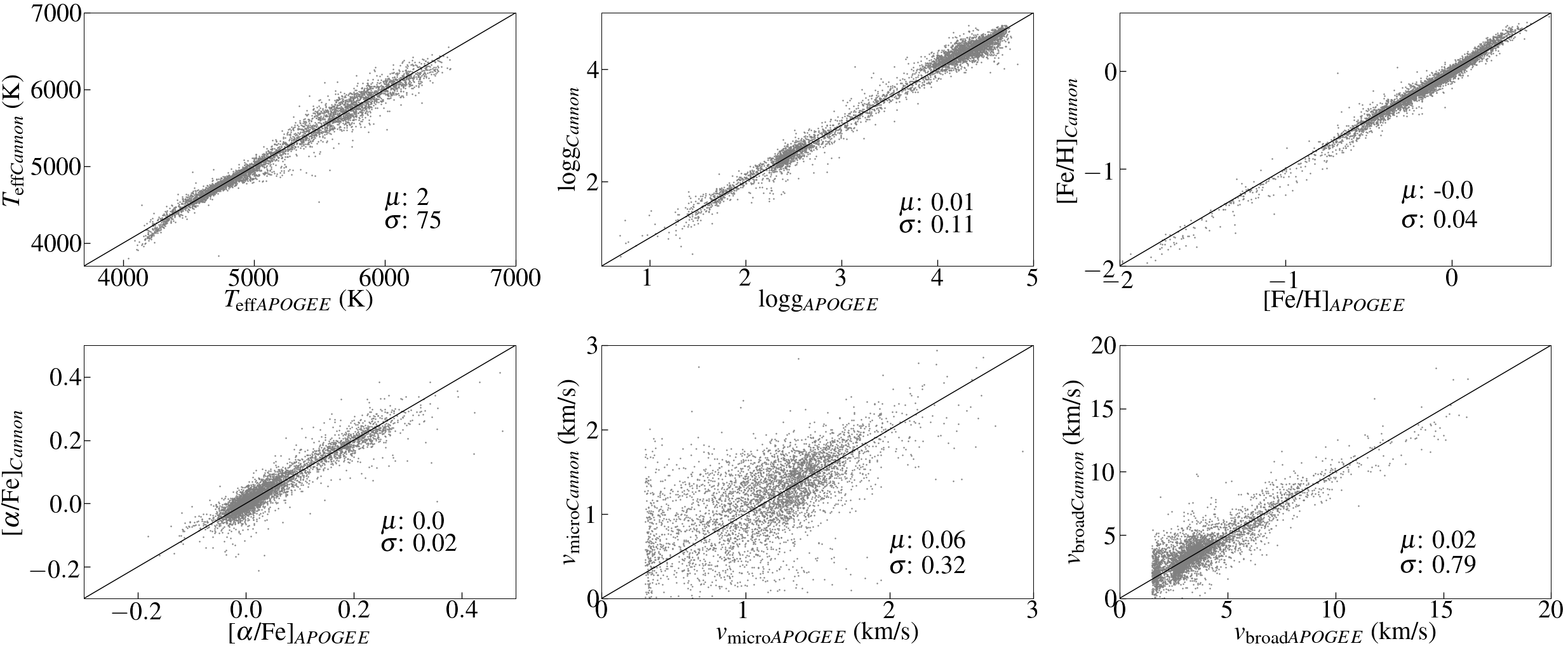}
    \caption{One-to-one relation of APOGEE ASPCAP stellar labels (x-axis) in the training set versus corresponding values estimated by the \textit{Cannon} from APOGEE spectra (y-axis) after carrying out a 12-fold cross-validation test. The mean and scatter (calculated as the mid value of 84$^{th}$-16$^{th}$ percentile) in each label is indicated on the bottom right hand side of each plot.   }
    \label{fig:CV_GalahApg_APOGEE}
\end{figure*}

We follow the method described in section~\ref{sec:Cross-Validation} to carry out 12-fold cross-validation of the \textit{Cannon} estimates for 6 labels using the training set. In the case of \textit{GALAH Cannon GALAH SME}, we compare the \textit{Cannon} outputs with input GALAH SME labels, while we compare the \textit{Cannon} outputs with input APOGEE ASPCAP labels for \textit{APOGEE Cannon APOGEE ASPCAP}. Figures~\ref{fig:CV_ApgGalah_GALAH} and ~\ref{fig:CV_GalahApg_APOGEE} show the cross-validation results for \textit{GALAH Cannon GALAH SME} and \textit{APOGEE Cannon APOGEE ASPCAP} respectively

\begin{table}
\caption{Table schema of APOGEE scaled and GALAH scaled stellar parameter catalogues}\label{table:scheme}
\begin{tabular}{l c c  }
\hline
\hline
Column Name & Unit  & Description   \\
\hline
Unique ID  &   & Unique survey star id   \\
dr2$\_$source$\_$id  &   & \textit{Gaia} dr2 ID   \\
dr3$\_$source$\_$id  &   & \textit{Gaia} dr3 ID   \\
ra  &  deg  &  Right ascension\\
dec  &  deg  &  Declination\\
teffCann &  K  & \textit{Cannon} estimate of effective temperature\\
e$\_$teffCann &  K  & error for teffCann\\
loggCann &  log(cm/s$^{2}$)  &  \textit{Cannon} estimate of surface gravity \\
e$\_$loggCann &  log(cm/s$^{2}$)  &  error for loggCann \\
fehCann &  dex  &  \textit{Cannon} estimate of metallicity \\
e$\_$fehCann &  dex  &  error for fehCann \\
alphafeCann &  dex  &  \textit{Cannon} estimate of [$\alpha$/Fe] \\
e$\_$alphafeCann &  dex  &  error for alphafeCann \\
vmicroCann &  km/s  &  \textit{Cannon} estimate of microturbulence \\
e$\_$vmicroCann &  km/s  &  error for vmicroCann \\
vbroadCann &  km/s  &  \textit{Cannon} estimate of broadening velocity \\
e$\_$vbroadCann &  km/s  &  error for vbroadCann \\
r$\_$chi$\_$sq  &   &  reduced chi-square from \textit{Cannon}\\
snrCann  &   &  Signal-to-noise ratio from \textit{Cannon}\\
Cannon$\_$distance &  & Distance to the training set labels\\
flag$\_$id &  & 0 - GALAH spectra, 1 - APOGEE spectra \\
flag$\_$Cannon$\_$dist &  & flag for quality of \textit{Cannon} estimates (0/1)\\
flag$\_$spectra &  & flag for problematic spectra  \\
flag$\_$sp$\_$aspcap &  & survey pipeline flag  \\
flag$\_$survey &  & invalid survey pipeline estimates (1) \\
flag$\_$training &  & star belonging to the training set (0) \\ 
teff  &  K  &  Survey effective temperature\\
e$\_$teff  &  K  &  Survey effective temperature error\\
logg  &  log(cm/s$^{2}$)  &  Survey surface gravity\\
e$\_$logg  &  log(cm/s$^{2}$)  &  Survey surface gravity error\\
feh  &  dex  &  Survey metallicity \\
e$\_$feh &  dex  &  Survey metallicity error \\
alphafe &  dex  &  Survey [$\alpha$/Fe] \\
e$\_$alphafe &  dex  &  Survey [$\alpha$/Fe] error \\
vmicro &  km/s  &  Survey microturbulence \\
e$\_$vmicro &  km/s  &  Survey microturbulence error \\
vbroad &  km/s  &  Survey broadening velocity \\ 
e$\_$vbroad &  km/s  &  Survey broadening velocity error\\
snr  &   &  Survey signal-to-noise ratio\\
r$\_est$  &  pc  &  Estimated distance from Bailer-Jones+21\\
r$\_lo$  &  pc  &  Lower bound from Bailer-Jones+21\\
r$\_hi$  &  pc  &  Higher bound from Bailer-Jones+21\\
parallax  &  mas  &  Gaia EDR3 parallax\\
e$\_$parallax  &  mas  &  Gaia EDR3 parallax error\\
\hline
\hline
\end{tabular}
\end{table}


In the case of \textit{GALAH Cannon GALAH SME}, all labels follow a tight one-to-one relation as indicated by the mean and scatter values as well. Largest scatter (0.48 km/s) is seen in the case of $v_{\rm broad}$, which is still much lower than in the case of \textit{APOGEE Cannon GALAH SME}.

In the case of \textit{APOGEE Cannon APOGEE ASPCAP}, all labels except $v_{\rm micro}$ and $v_{\rm broad}$ follow a tight one-to-one relation. Still, we find the \textit{Cannon} to under estimate T$_\mathrm{eff}$ for cool stars (T$_\mathrm{eff}$$_{\rm, APOGEE}$ $<$ 4500 K) and stars with T$_\mathrm{eff}$$_{\rm, APOGEE}$ ~ 5000 - 5400 K, while slightly over estimate T$_\mathrm{eff}$ for stars with T$_\mathrm{eff}$$_{\rm, APOGEE}$ in between these two limits. Similarly, we find the \textit{Cannon} to slightly over predict $\log g$ for stars with $\log g$$_{\rm APOGEE}$ $>$ 4 dex. 

There are significant deviations in the case of $v_{\rm micro}$ and $v_{\rm broad}$, similar to that seen in the case of \textit{GALAH Cannon APOGEE ASPCAP} (see Figure~\ref{fig:CV_GalahApg}). We find the \textit{Cannon} outputs for input $v_{\rm micro}$$_{\rm APOGEE}$ $>$ 1 km/s follow one-to-one relation, whereas the \textit{Cannon} estimates have higher dispersion for lower $v_{\rm micro}$$_{\rm APOGEE}$ values. This indicates that the \textit{Cannon} model is unable to find significant correlation between $v_{\rm micro}$$_{\rm APOGEE}$ values and corresponding APOGEE spectra. Much tighter one-to-one trend is seen in the case of $v_{\rm broad}$, though there are significant deviations for lower $v_{\rm broad}$$_{\rm APOGEE}$ values. As mentioned in Section~\ref{sec:Cross-Validation}, this again possibly indicates the inefficiency/problems associated with the way $v_{\rm micro}$ and $v_{\rm broad}$ have been determined for APOGEE stars.

\begin{figure*}
	\includegraphics[width=\textwidth]{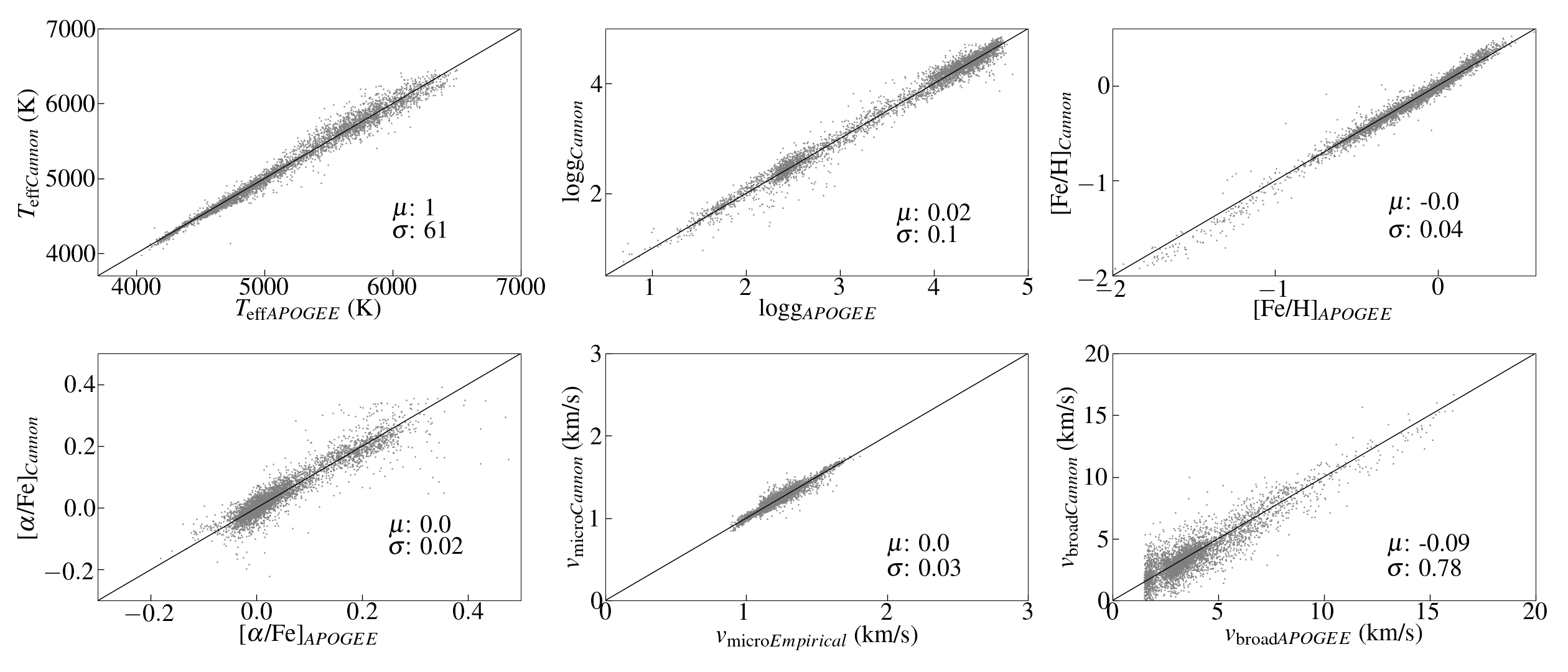}
    \caption{One-to-one relation of APOGEE ASPCAP stellar labels (x-axis) in the training set versus corresponding values estimated by the \textit{Cannon} from APOGEE spectra (y-axis) after carrying out a 12-fold cross-validation test. The mean and scatter (calculated as the mid value of 84$^{th}$-16$^{th}$ percentile) in each label difference is indicated on the bottom right hand side of each plot.   }
    \label{fig:CV_GalahApg_APOGEE_emp}
\end{figure*}

\begin{figure*}
	\includegraphics[width=\textwidth]{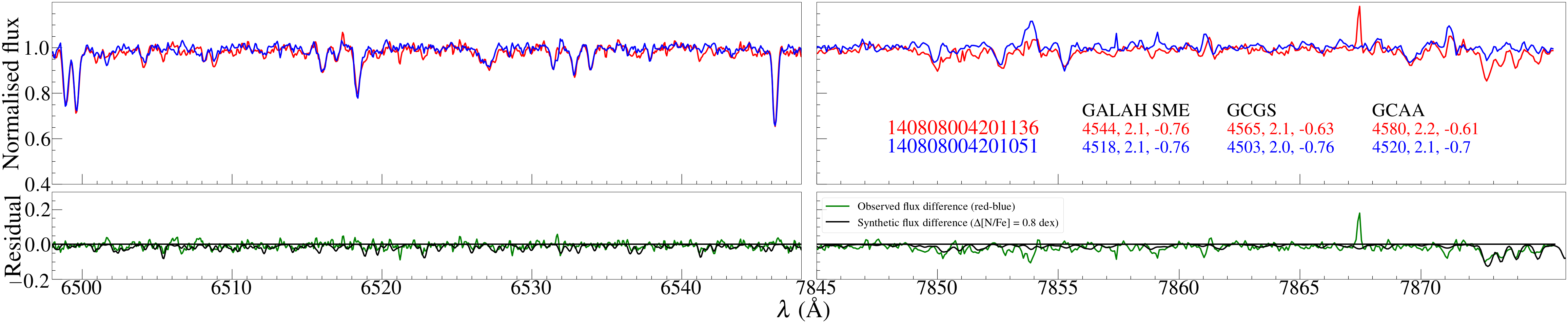}
    \caption{GALAH spectra for two NGC 104 members with similar stellar parameters from GALAH. In blue, we show star spectrum for which GALAH and APOGEE scaled stellar parameters from GALAH spectra (GCGC and GCAA) are consistent with GALAH SME. Star spectrum in red is an example where GCGS and GCAA metallicity estimates are more metal rich than GALAH SME value. This inconsistency can be attributed to the difference between the two spectra(Red-blue), shown in green in the bottom panels (mean flux difference is indicated in these panels), arising most likely due to wrong continuum normalisation for red spectra. }
    \label{fig:NGC104_cont}
\end{figure*}

\begin{figure*}
	\includegraphics[width=\textwidth]{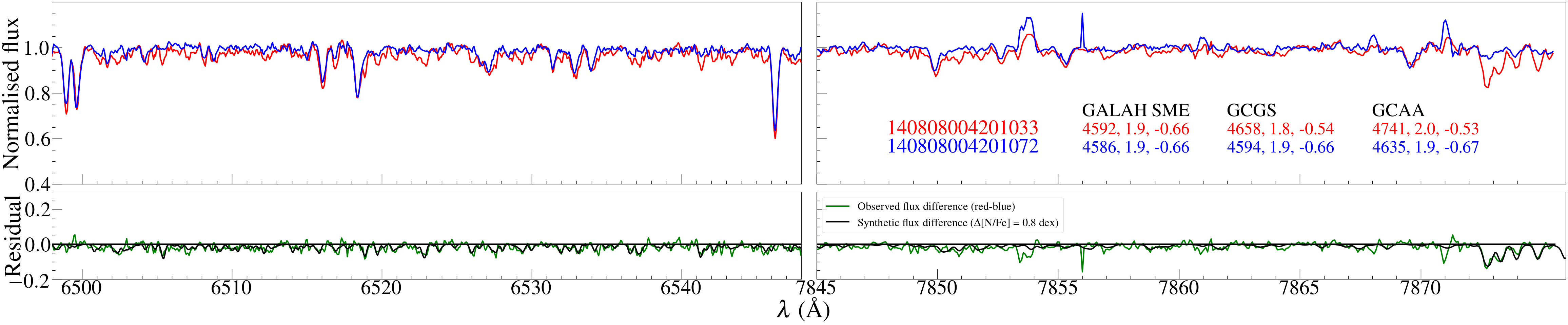}
    \caption{Same as Figure~\ref{fig:NGC104_cont} but for another pair of stars with similar stellar parameters in NGC 104.  }
    \label{fig:NGC104_strength}
\end{figure*}

To investigate this further, we carry out an exercise in which instead of the APOGEE $v_{\rm micro}$, we make use of the empirical relation adopted in GALAH (Equations 4 and 5 in \citealt{GALAHDR3}) to determine $v_{\rm micro}$ for APOGEE as well. We do this since the \textit{Cannon} could correlate the $v_{\rm micro}$ from GALAH SME with APOGEE spectra in \textit{APOGEE Cannon GALAH SME} (see Figure~\ref{fig:CV_ApgGalah}). Including this new $v_{\rm micro}$, we train \textit{Cannon} model on APOGEE spectra using APOGEE labels. The cross-validation results from this exercise are shown in the Figure~\ref{fig:CV_GalahApg_APOGEE_emp}. First of all, there is a tight one-to-one relation for $v_{\rm micro}$. This is no surprise since the new empirical $v_{\rm micro}$ is a function of T$_\mathrm{eff}$ and $\log g$, and thus the \textit{Cannon} could identify tight correlation with corresponding spectra as well. This is again reflected in the T$_\mathrm{eff}$ and $\log g$ estimated by the \textit{Cannon}, which no longer show the issues that we found when APOGEE $v_{\rm micro}$ is used to train the \textit{Cannon} model. This can be considered as an evidence of the inefficiency/problems associated with the way $v_{\rm micro}$ have been determined for APOGEE stars.

\section{Metallicity difference in NGC 104}

As mentioned in section~\ref{sec:gc}, GCGS and GCAA metallicity estimates are found to be more metal rich than both survey pipeline estimates as well as \textit{Cannon} estimates from APOGEE spectra. To understand the reason for this, we compare GALAH spectra of two stars in NGC 104 with similar stellar parameters, out of which GCGS and GCAA metallicity is the same as GALAH SME estimate for one star and not for the other. Two such examples in the red and infrared wavelength ranges are shown in Figures~\ref{fig:NGC104_cont} and \ref{fig:NGC104_strength}, where spectra in blue are stars with consistent \textit{Cannon} estimate and red spectra represent the inconsistent case. Residuals resulting from difference between red and blue spectra, is shown in green in the bottom panels. We have computed illustrative synthetic spectra following \cite{GALAHDR3} for the relevant stellar parameters with a solar chemical composition as well as with an enhancement in N of 0.8 dex, the difference between which is plotted in black in the same panel. The residual from observed spectra clearly follow that from the synthetic spectra, especially evident from the CN bands at $>\sim$ 7870\,\AA, suggesting that the stars with more metal rich \textit{Cannon} estimates (red spectra) have enhanced N abundances and thus stronger CN bands.




\end{appendix}



\label{lastpage}
\end{document}